%% file: main.tex
\documentclass[MS, 12pt]{ucdavisthesis}

\usepackage{etoolbox}
\makeatletter
\patchcmd{\@footnotetext}{\footnotesize}{\scriptsize}{}{}
\makeatother

\usepackage[us,nodayofweek,12hr]{datetime}
\usepackage{graphicx}
\usepackage[round]{natbib}
\usepackage{hyperref}
\usepackage{float}
\usepackage{color,soul}
\usepackage{amsmath}
\usepackage{booktabs}
\usepackage{pbox}
\usepackage{amssymb}
\usepackage{comment}
\usepackage{wrapfig}
\usepackage{caption}
\usepackage{subcaption}
\usepackage{array}
\usepackage{multirow}
\usepackage{tabu}
\usepackage[utf8]{inputenc}
\usepackage{csquotes}
\usepackage{textgreek}

\title          {NIRS Based Bladder Volume Sensing for Patients Suffering with Neurogenic Bladder Dysfunction}
\author         {Prashant Gupta}
\authordegrees  {B.Tech. (Maharshi Dayanand University, India) 2015}
\thesis{Master of Science}

\officialmajor  {Electrical and Computer Engineering}
\graduateprogram{Electrical and Computer Engineering}
\degreeyear     {2018}
\degreemonth    {June}
\committee{Professor Soheil Ghiasi}{Professor Hussain Al-Asaad}{Professor Venkatesh Akella}{}

\nocopyright

\dedication{\textsl{Dedicated to my Parents, \\ without whom this would have been still a dream ...}}

\abstract{
Neurogenic Bladder Dysfunction has detrimental effects on day-to-day life of millions of people. Some of the most common symptoms faced by these patients include urinary incontinence, urgency and retention. Since elevated bladder pressure due to prolonged urine storage inside bladder may have adverse impacts on patient's renal health, urologists recommend clean-intermittent catheterization (CIC) every 2 to 4 hours throughout the day to relieve bladder pressure. However, since urine production by kidneys is an intermittent process and most of these patients have limited mobility, such frequent trips to washroom can prove to be challenging. Sometimes, bladder fills to capacity before the recommended CIC time is reached causing embarrassing situation due to leakage. Hence, time-based CIC strategy is difficult to implement and has high chances of failure. As such, continence is the primary concern for most of these patients but sadly there are no practical solutions available in the market that address this concern.

A real-time notification system that could give feedback to patients on when \enquote{bladder is almost-full} could help these patients to better plan their bathroom trips. This work explores the feasibility of using a near infrared-light based wearable, non-invasive spectroscopy technique that can sense amount of urine present inside the bladder and give details on developing a bladder state estimation device.

We present preliminary results by testing our device on optical phantoms and performing \textit{ex vivo} measurements on porcine bladder and intestines. We later explored the possibility of using the device on human subjects, after study was approved by the UC Davis Institution Review Board (IRB).
}

\acknowledgments{
First and foremost I would like to express my sincere gratitude to my advisor Professor Soheil Ghiasi, for mentoring me through these two wonderful years and always steering me in the right direction. This thesis wouldn't have been possible without his constant and able guidance. I consider myself privileged to work with him and thank him for the time he has invested in me. I would also like to take this opportunity to thank my committee members Professor Hussain Al-Asaad and Professor Venkatesh Akella for taking the time to review my work and giving insightful comments. Their feedback helped me towards improving this thesis.

Second, I would like to thank my fellow graduate students Daniel Fong, Alejandro Velazquez, Mahya Saffarpour, Rami Abueshsheikh at UC Davis who played a crucial part in this study. Parts of this thesis are based upon collaborative work done with them and I consider myself lucky to be working in conjunction with such great minds. Discussions I had with Mohammad Motamedi and Terry O'Neil, two humble and truly bright PhD students, helped me enrich my knowledge on a variety of technical domains apart from my research. I am grateful to Ms. Seema Bansal for proof reading my draft, helping me with some of the figures included in this thesis and most of all believing in me.

Third, I would like to thank my friends at UC Davis Praveen, Ismail, Ajinkya, Saheel, Sachin, Sandeep, Satyabrata, Jennifer, Shweta and Deepika who supported me during my studies and whose fair share of humour always kept me going through stressful times. I look forward to our continued friendship and wish them well for their endeavours.

Finally, I would like to express my eternal debt to my parents and elder sister Neha, who always inspire me to pursue my dreams and work hard towards achieving them. A lot of this work is based on constant trial and error and its their continuous love, support and belief in me that kept me motivated through my ups and down.

A big thank you is owed to the volunteers who participated in this study; without their enthusiastic support, this research project would not have been possible. You are the true champions of science.
}

\begin{document}

%\newcommand{\bibfont}{\singlespacing}
% need this command to keep single spacing in the bibliography when using natbib

\setcounter{tocdepth}{2}
\bibliographystyle{unsrtnat}

%many other bibliography styles are available (IEEEtran, mla, etc.). Use one appropriate for your field.

\makeintropages %Processes/produces the preliminary pages

\include{Chapter1}
\include{Chapter2}
\include{Chapter3}

\include{Chapter4}
\include{Chapter5}
\end{document}

%% file: Chapter1.tex
\chapter{Introduction}
\label{chapter1}

Human bladder is a hollow balloon shaped muscular organ which is a critical part of the urinary track system and stores urine until the person finds an appropriate time and place to urinate. According to Urology Care Foundation, more than 33 million people suffer from neurogenic bladder dysfunction (in US alone). This is a well-documented problem in which nerves carrying the messages between bladder, spinal cord and brain don't work congruously and patients lack sensation and control of their bladder. People affected from Spinal Cord Injuries (SCI), congenital spinal anomalies like spina bifida, injuries like herniated discs, men post prostate cancer removal, women post childbirth or menopause commonly suffer from this problem  \citep{white2016spinal, broome2003impact}. Some of the most common symptoms faced by these patients include urinary incontinence, urgency and retention. Recently, several studies have also shown prevalence of stress related incontinence in adolescents or young adults \citep{robinson2014urinary}. Unawareness of bladder filling may lead to development of high pressure within the urinary tract and can severely damage kidneys.

While prevention of renal failure is of paramount importance, primary day-to-day concern of most patients is incontinence. The commonly used solution in such cases is Clean Intermittent Catheterization (CIC) which is recommended every 2 to 4 hours to prevent any leakage and does not involve any surgery or permanent appliance attachment. The procedure enables drainage of urine by inserting a catheter, a thin hollow tube, into bladder to help relieve bladder pressure. If not done carefully, it can often lead to infection, urethral erosion and other complications \citep{gray1995incontinence}. As urine production by kidneys is not a constant process, it becomes difficult to predict bladder filling using time-based techniques.

 With limited mobility in most cases, one common problem reported by these patients is, making a difficult trip to bathroom only to find limited amount of urine in the bladder. This leads to unnecessary catheterization which can further damage the bladder. Or worse, not making it to the bathroom in time and leakage of urine because of full bladder. To prevent such accidents, particularly in public, these patients have to carry absorbent products, spare clothing and organize fluid intake. Some patients go so far as to choose a permanent indwelling catheter which may lead to higher chances of chronic infection and bladder cancer \citep{nahm2015bladder}. Such situations can severely impact the self-efficacy of these patients causing barriers to social, recreational activities and in some cases may even lead to depression \citep{broome2003impact}. Quality of life of these patients could dramatically improve if they have knowledge of the right time to void. That way these patients could plan their trips to washroom in advance, avoid incontinence, leaking and in-turn have a better quality of life. 

Although a variety of tools have been developed to measure amount of urine inside the bladder, nearly all these devices are costly, big in size and are made with focus on clinicians/caregivers rather than patients. Usually a specially trained person is required to interpret these results and communicate them to patient. 

%TODO: Add more details on what this thesis is about. 
This thesis talks about steps towards development of a wearable non-invasive device for monitoring changes in bladder volume by optical monitoring using Near-Infrared Spectroscopy (NIRS). Goal of this research is to develop a device that can communicate directly with patients to give them real-time feedback on the right time to void and at the same time cause minimum interference in their day-to-day activities. Such a device can thus enable these patients to empty their bladder when they actually have to and thus, bring back some normalcy to their lives.

Work done in this study is supported by CITRIS and the Banatao Institute at the University of California. Parts of this thesis are based on two previous publications with other authors:

\begin{itemize}

\item \textbf{Non-Invasive Bladder Volume Sensing for Neurogenic Bladder Dysfunction Management}. Daniel Fong, Alejandro Velazquez Alcantar, \textbf{\textit{Prashant Gupta}}, Eric Kurzrock, and Soheil Ghiasi presented at 15$^{\text{th}}$ IEEE International Conference on Wearable and Implantable Body Sensor Networks (BSN), March 2018, Las Vegas, USA \citep{dfong2018lepsbv}. 

\item \textbf{Restoring the Sense of Bladder Fullness for Spinal Cord Injury Patients}. Daniel Fong, Xiaofan Yu, Jiageng Mao, Mahya Saffarpour, \textbf{\textit{Prashant Gupta}}, Rami Abueshsheikh, Alejandro Velazquez Alcantar, Eric Kurzrock and Soheil Ghiasi accepted in IEEE/ACM 3$^{\text{rd}}$ Conference on Connected Health: Applications, Systems and Engineering Technologies, September 2018, Washington, D.C. \citep{df2018lepsbv2}.

\end{itemize}

%% file: Chapter2.tex
\chapter{Background}
\label{chapter2}
This chapter covers relevant background theory and provides an overview to the research that has previously been conducted in the field of bladder volume measurement. As this study is an intersection of various domains, the idea behind this section is to give a high-level overview of the optical and biological parts involved in this research. Details on embedded systems and phantom trials are covered in Chapter \ref{chapter3}, while results on human trials are covered later under Chapter \ref{chapter4}.
%bio-technology, optics and embedded systems
%This chapter covers the relevant background theory, current research trends in the fields of NIRS, bladder volume measurement and tries to highlight the need for a study like this.  

\section{Near Infrared Spectroscopy (NIRS)}
NIRS is an optical measurement technique that uses near-infrared region (700nm to 2000nm) of electromagnetic spectrum to investigate  tissue composition. J\"obsis in 1977 first demonstrated its use by measuring change in concentration of oxy- and deoxy- hemoglobin for brain and muscle tissues \citep{jobsis1977noninvasive}. This technique became widely popular because of its characteristics like  non-invasiveness, affordability and ease-of-use. Some of its current applications include astronomy, agriculture, remote monitoring, material science and a large number of medical applications like pulse oximetry \citep{scheeren2012monitoring, fong2017transabdominal}, brain computer interface, rehabilitation etc. Recently some researchers have shown potential of using NIRS for urology applications like measuring changes in bladder content \citep{molavi2014noninvasive}, bladder contractions, tissue response to physiological events etc. 

\subsection{Principle}

Optode for NIRS consists of two major components - emitter and detector. While emitter could be any light source from halogen light bulbs to LEDs, the main criterion of detector selection is that it should be sensitive to the wavelength of photons as emitted by the emitter optode. Another trade-off worth noting is that, as the size of detectors increases, the probability of detecting photons also increase, at the same time it becomes more susceptible to background noise. On the other hand as detector size decreases probability of detecting photons also decreases and at the same time it is less susceptible to background noise.

NIRS is usually performed by emitting light towards the tissue surface and measuring diffused-reflected light that escapes from the surface after it has travelled some distance from the emitter. While the visible light only penetrates human tissue for short distances since it is markedly attenuated by several tissue components, the near infrared (NIR) spectrum photons are capable for deeper penetration (upto several centimeters or more). In continuous NIRS trend monitoring, a constant optode - skin contact is very important, as minor changes in contact induces major changes in the signal value. 

\begin{figure}[ht]
\centering
\includegraphics[width=6in]{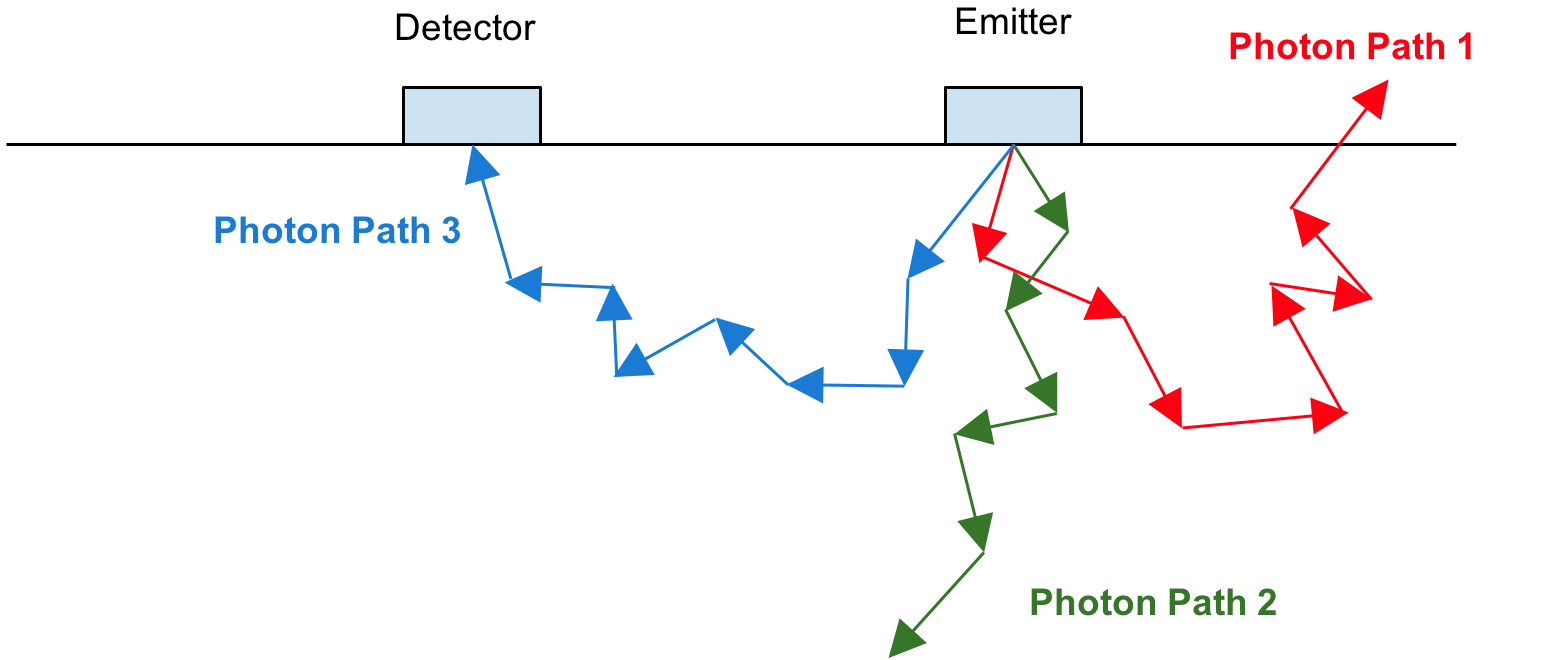}
\caption[Potential photon paths]{Photon path 1, 2 and 3 depict potential paths that a given photon can travel. After multiple reflections through the biological tissue; photon 3 is successfully detected at detector, photon 2 ends up getting absorbed by tissue, while photon 1 escapes the tissue structure and is neither absorbed by tissue nor detected by detector.}
\label{figure1}
\end{figure}

Propagation of light through the biological tissue is governed by three major phenomenon namely reflection, absorption and scattering. Angle with which light enters the tissue (with respect to tissue surface) determines light reflection and in-turn decides the path through which photons travels within the medium.  Photons entering the body through emitter goes through multiple reflections and can potentially have three possible outcomes as described in Figure \ref{figure1}. Photon 1 after multiple reflections escapes body without being detected, Photon 2 after multiple reflections gets absorbed by tissue while Photon 3 ends up successfully getting detected by detector. Out of the total near-infrared light entering the system, approximately 80\% of the attenuation is as a result of scattering while the remaining 20\% is lost due to absorption. Hence, one of the biggest hurdle while attempting to do quantitative measurements with NIRS is loss due to scattering.

\subsection{Modes of Operation}
\label{modeofOperation}
NIRS has two main modes of operation, namely- \textit{Transmission Mode} and \textit{Reflection Mode}.

In transmission mode, emitter is placed on one side of the biological tissue while detector is placed on the other end. Photons in this mode travel through the entire tissue structure and therefore, can get global information of the tissue medium. Transmission mode is usually suitable when thickness of the medium is less than 8 cm \citep{pellicer2011near}. For example, it is popularly used in infants to get information about their brain's oxygenation.

\begin{figure}[ht]
\centering
\includegraphics[width=5in]{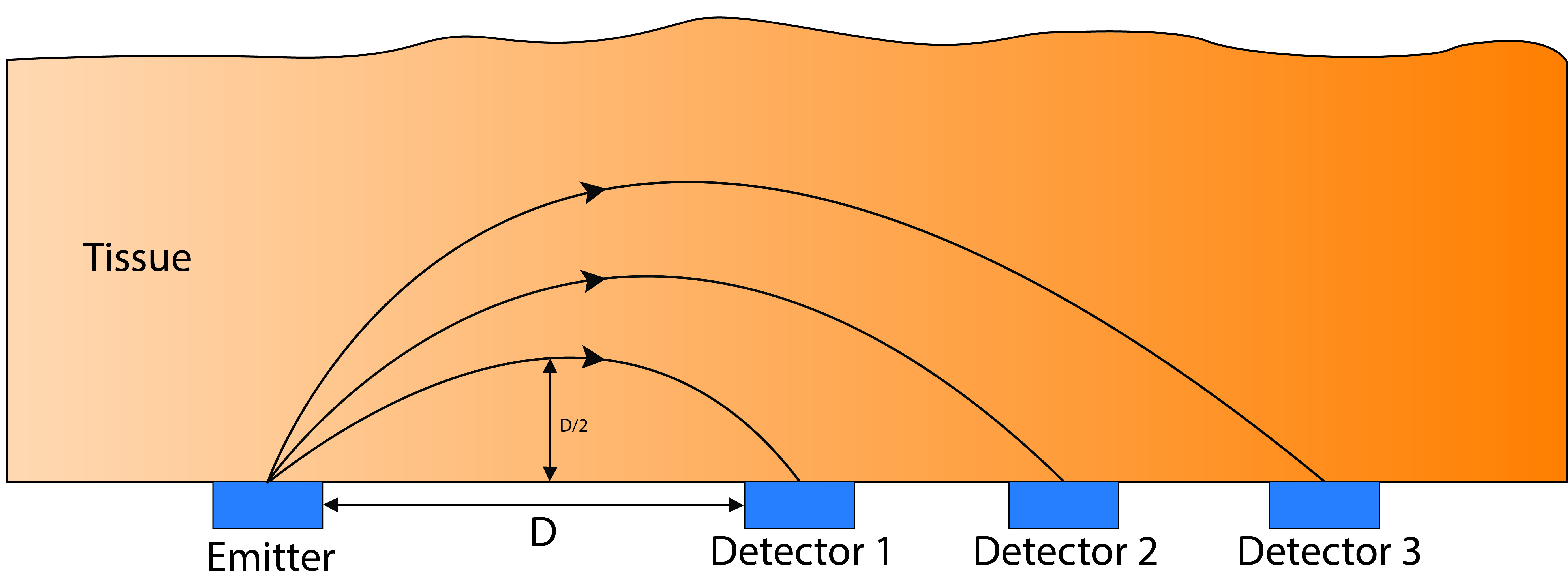}
\caption[Reflection Mode Operation NIRS]{Optode pairs working in reflection mode are placed on the same side of optical tissue. Photon path in this case can be approximated to follow a curved trajectory having a penetration depth of \enquote{D/2} where, \enquote{D} is the emitter (source) - detector distance. As detector moves farther from emitter, intensity of detected light decreases, while penetration depth of photon increases.}
\label{figure2}
\end{figure}

In reflection mode, both emitter and detector are placed on same side of the biological tissue and therefore, can only extract regional information as photons only travel though a shallow depth. Light in such a case can be approximated to follow a curved trajectory as depicted in Figure \ref{figure2} and the maximum depth of that trajectory is approximately half of the source-detector (SD) distance. As distance between detector and emitter increases, penetration depth of photon entering the body also increases; at the same time, number of photons reaching detector decreases because of increased path-length resulting in higher attenuation due to increased probability of tissue absorption. In this study, the developed device is operated in reflection mode so that the photons can successfully reach the human bladder (2-5cm deep from the skin surface) after passing though multiple layers of tissue i.e. muscle, fat etc.

\subsection{Measurement of Optical Signal}
% TODO: fill this up
Detection of optical signal coming out from the tissue structure post diffused-reflectance requires 3 major components- Optical detector (e.g. Photodiode), Optode actuation subsystem, Data handling and control subsystem. 

Optical detector measures optical signal and converts it into electrical current using photoelectric effect. For detectors to measure optical signal as a result of emitter optode, emitter-detector optode pairs have to actuate in synchronization. Also, since the detected optical signal is prone to noise as a result of variable ambient light, an ambient light measurement is also taken at detector while emitter is turned off. This ambient light signal can later be subtracted from the detected signal (while emitter is turned-on) to remove the impact of noise due to ambient light. 

\begin{figure}[h]
\centering
\includegraphics[width=6in]{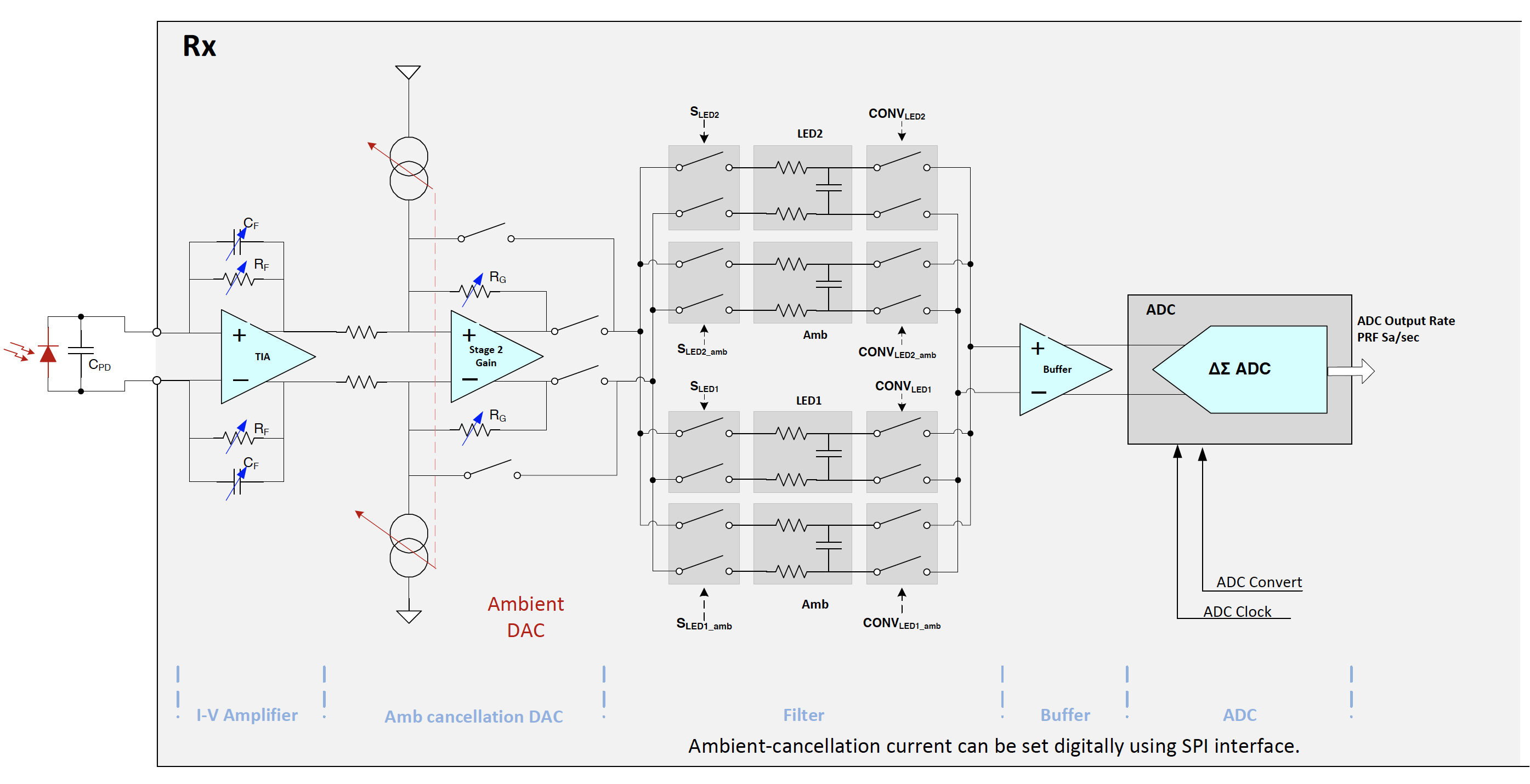}
\caption[Schematic diagram of Receiver-Front End for TI AFE4490]{Schematic diagram of Receiver-Front End for TI AFE4490 \citep{ti:afe4490}.}
\label{fig:afe4490}
\end{figure}

Many electrical measurement components use voltage signals rather than current signals to buffer, convey and manipulate information because of low transmission loss, low power consumption, design flexibility etc. Hence, the current signal measured from the optical detector is thus converted into voltage signal using a transimpedence amplifier (TIA).  Figure \ref{fig:afe4490} shows the receiver-front of TI AFE4490 that has been used for optode actuation and data handling in this thesis. Gain of TIA can be controlled by changing the resistance ($R_F$) across operational amplifier (op-amp) as seen by the I-V Amplifier stage of the AFE.  AFE4490 also provides a user-controlled ambient light cancellation system however, the ambient cancellation can also be achieved in post-processing stages. Lastly, the voltage signal present in analog domain is converted to digital domain using sample-and-hold circuitry present in the analog-to-digital converter (ADC). Conversion of analog signal into digital domain comes with various advantages like less expensive signal processing, less susceptible to noise etc. The digital signal obtained out of ADC can thus be used to investigate the optical properties of tissue structure. 

\subsection{Beer-Lambert Law(BLL)}

BLL gives out a relationship between light attenuation and properties of the medium it is travelling through. The law is commonly applied to chemical analysis measurements like analysis of a mixture by spectrophotometry and does not require any extensive sample pre-processing. Investigating tissue composition from NIRS-based measurements is accomplished through a modified version of the Beer-Lambart Law (MBLL):

\[A(\lambda) = log (I_O/I) = \sum \epsilon_i c_i D L + G \]

\begin{figure}[ht]
\centering
\includegraphics[width=6in]{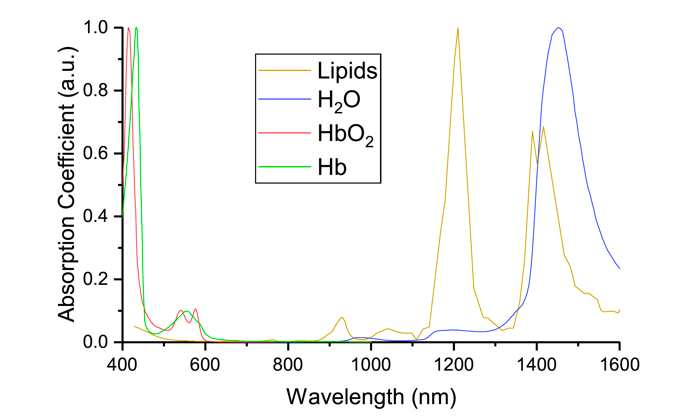}
\caption[Absorption Coefficient of main chromophores in human tissue]{Normalized absorption coefficient spectra for the main chromophores
in tissue, namely water \citep{kou1993refractive}, oxy- and deoxy- hemoglobin \citep{prahl1999tabulated},
and lipids \citep{van2004determination}, \citep{wilson2015review}. Image source- \citep{dfong2018lepsbv}.}
\label{figure3}
\end{figure}

Namely, the MBLL says that absorption of light in tissue $A(\lambda )$, is log of the ratio of  incident light intensity $I_o$, to transmitted light intensity $I$. More usefully, it says that absorption of light is a function of the concentration of chromophores in tissue $c_i$, their molar extinction coefficients $\epsilon_i$, source-detector distance $L$, differential path-length factor $D$ and a term that accounts for the static attenuation due to light scattering $G$. The molar extinction coefficient $\epsilon$ (units: $cm^{-1}$/(moles per liter)) represents level of light absorption experienced and is sometimes represented as absorption coefficient $\mu_a$ (units: $cm^{-1}$). Both coefficients represent a chromophore's level of absorption per concentration and per unit length, and differ only by a scaling factor. Similarly, the scattering coefficient $\mu_s$ (units: $cm^{-1}$) represents expected number of scattering events per unit length and is inherently a part of both $D$ and $G$. In human tissue, some of the main chromophores include oxy- and deoxy- hemoglobin ($HbO_2$ and $Hb$), lipids (fat), melanin (skin pigment), and water. Their normalized absorption coefficients, which is related to molar extinction coefficients, can be seen in Figure \ref{figure3}.

\section{Human Bladder}
\begin{figure}[h]
\centering
\includegraphics[width=6.2in]{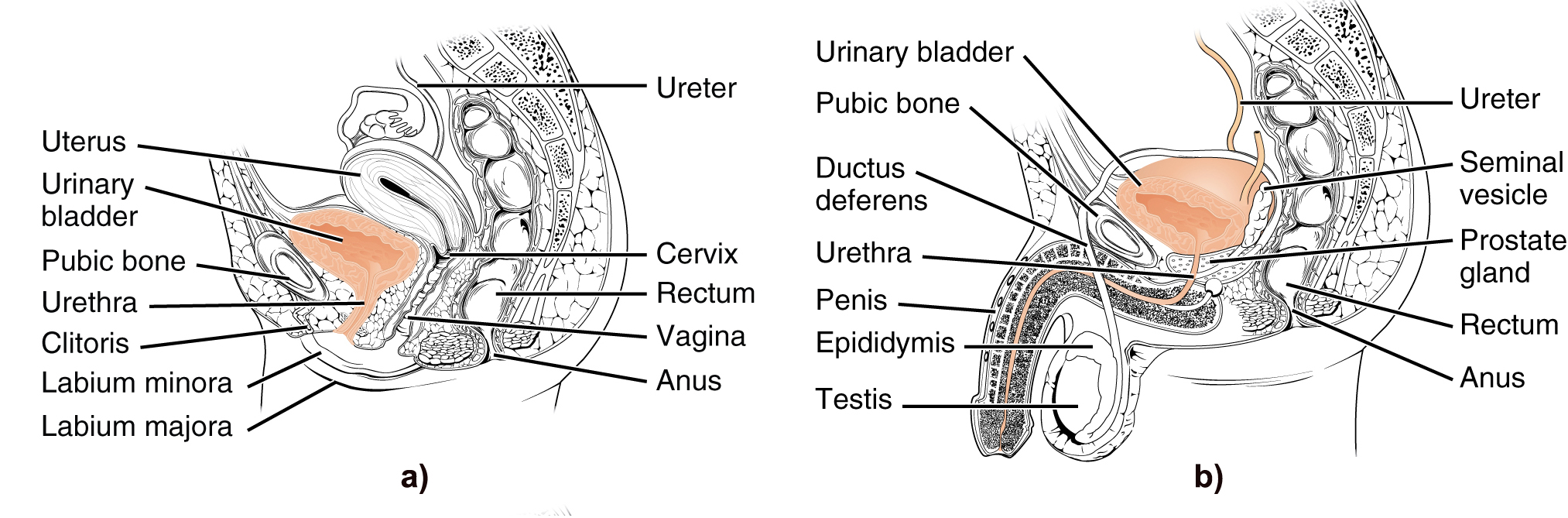}
\caption[Female and Male Bladder]{Drawing of a)Female Bladder b) Male Bladder \citep{openstax}}
\label{figure5}
\end{figure}

In NIRS, path of light and medium through which light propagates has a significant impact on results and hence it becomes critical to study basic anatomy and physiology of human bladder. Figure \ref{figure5} shows a drawing depicting female and male bladder.

\subsection{Anatomy}
Bladder is a muscular sac, located in lower abdomen of human body just above and behind the pubic bone, which collects urine produced by kidneys before being voided. While different scientists have different theories about human bladder capacity and how much urine it can hold, general consensus being a normal functioning capacity in adults ranges from 400 to 600ml. For a non-invasive application of NIRS photons would have to travel through multiple layers of tissue before reaching bladder (having a thickness of about 2-5cm). Different layers of tissues that come before bladder wall are namely- Outer Skin, Subcutaneous tissue, Adipose tissue, Fascia, Muscle and Retropubic space.

\subsection{Physiology}
\begin{figure}[h]
\centering
\includegraphics[width=4in]{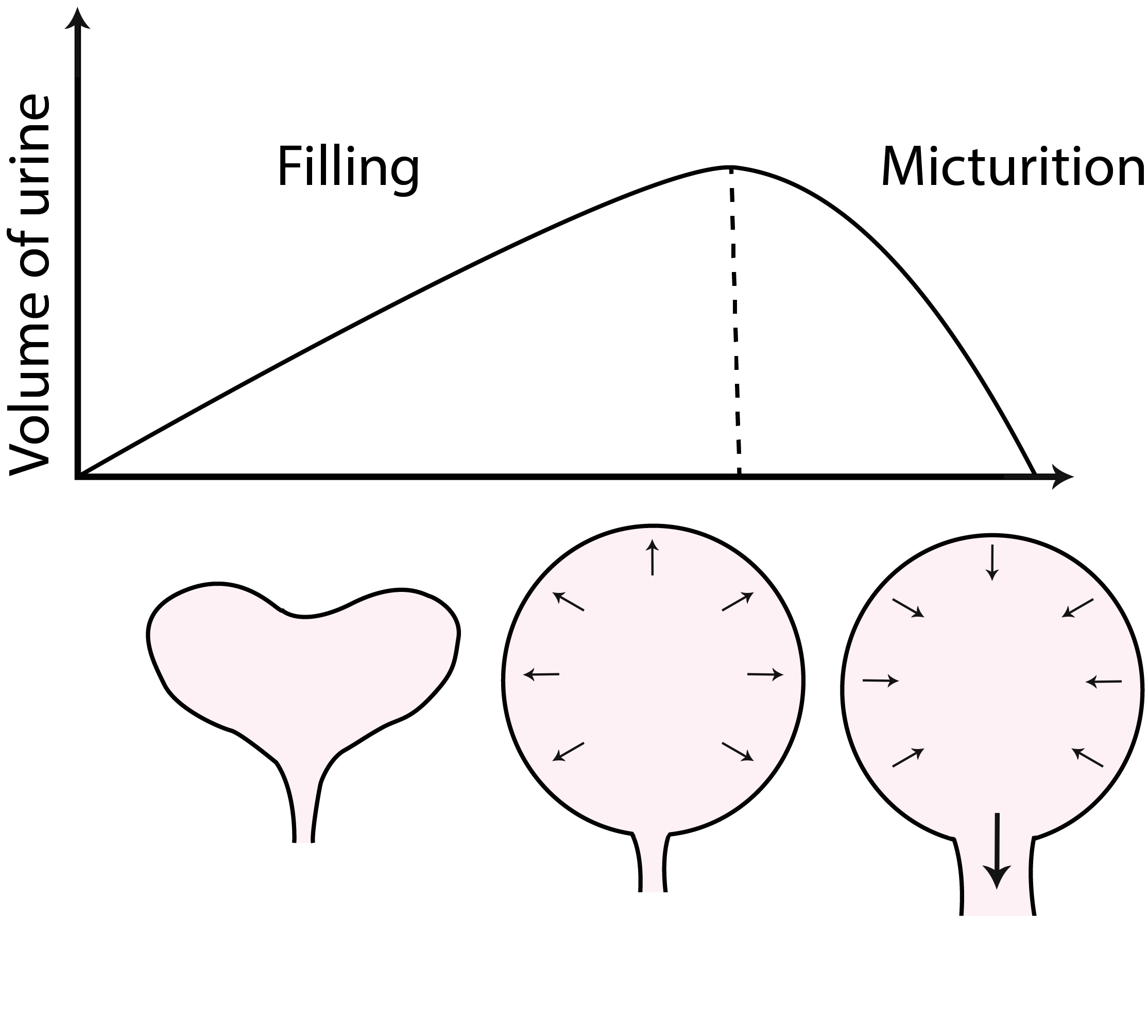}
\caption[Bladder shape change due to change in bladder volume]{Illustration depicting change in bladder shape as the volume of urine inside bladder changes.} 
\label{figure4}
\end{figure}

Kidneys filter blood flowing through the body and remove unwanted chemicals with excess water (91-96\% of urine \citep{rose2015characterization}) which all combine to form urine. After urine is produced by kidneys, it goes though ureters to reach bladder. As volume of urine inside bladder increases, muscles of bladder accommodate urine by passive relaxation controlled by the autonomic nervous system. Bladder starts to stretch like a balloon and starts getting more circular in shape \citep{kristiansen2004effect} to accommodate the urine being created. At the same time, thickness of bladder wall also decreases to less than 3mm (3-5mm when bladder is empty). Figure \ref{figure4}, shows change in shape of bladder as it starts to fill up. While bladder is filling up, first sensation happens at a volume of about 150ml, and first desire to void happens at volume $>$350ml. As it approaches a volume of $>$450ml desire to void keeps getting stronger \citep{patel2010imaging}. A healthy human body is usually able to accommodate upto 500ml of liquid with comfort without any abnormal detrusor pressure. While micturating, pontine portion of the brain (cerebrum) sends a signal to the spinal cord that initiates a cascade of nervous signals and reflexes that allow volitional voiding. Bladder contracts and almost simultaneously bladder neck and urethral sphincter open. This gives humans the ability to micturate infrequently and voluntarily. 

\section{Related Work}
A number of investigators in recent times have tried to explore possibilities of estimating the volume of bladder for people suffering with neurogenic bladder. \cite{dreher1972bladder} and \cite{wang2009design} used an implantable magnet to interact with an external electronics switch to sense changes in bladder shape and in turn predict its volume. Such an approach is prone to errors due to interference from earth's and other peripheral magnetic fields. In addition to that, the magnetic strength degrades over time and can further hamper patient's ability to have an MRI (Magnetic Resonance Imaging) scan post implantation. \cite{coosemans2005autonomous} and \cite{majerus2016wireless} attempted to develop an implantable pressure sensor to measure bladder pressure. The inherent elasticity of bladder allows it to fill close to its functional capacity without significant change in pressure. Even as bladder pressure increases, it's often very close to the leaking point which does not give enough time to patients suffering from SCI to find a washroom and perform CIC. Furthermore, urinary retention beyond this point may have negative side effects, particularly in SCI patients, such as renal damage caused by back pressure in the urinary tract. Another approach includes use of MEMs enabled implantable strain gauge sensors which detect change in bladder size to predict bladder volume \citep{chen2015design}. This comes with issues of power, bio-compatibility, telemetry etc. Bladder tissue attached to the probe is likely to develop fibrosis after sometime which will alter the physical properties of tissue like reduce its ability to stretch, making the technique ineffective for long term use.  \cite{schlebusch2014bladder} proposed idea of using electrical-impedance tomography to sense conductance distribution of pelvic region using a belt with multiple electrical contacts. The approach has not been very successful in practice due to unreliability of electrical contacts with skin and highly variable tissue composition causing variation in impedance. A group of researchers used 950nm NIRS device to measure changes in oxy- and deoxy- hemoglobin of bladder wall, in which most notable changes occur during voiding \citep{macnab2005clinical}. This technique was thus used to determine full versus empty states of bladder by measuring attenuation in light due to water \citep{molavi2014noninvasive}. However, SCI patients showed inconsistent trends in tissue oxygen saturation, likely due to neurogenic bladder. Furthermore, chromophore changes that occur during voiding are not helpful in warning patient when bladder volume increases.

Currently in clinics, Doppler Ultrasound Tomography is used to measure bladder dimensions and using that information volume of bladder is thus predicted assuming it having an elliptical shape \citep{kiely1987measurement}. This device is bulky, expensive and has high error rate of upto 25\% \citep{dicuio2005measurements}. Recently, portable and wearable ultrasonic devices
for bladder monitoring have been developed but they are either
inaccurate or too computationally intensive to house the computational unit within the wearable part of the sensor \citep{kristiansen2004design}. In addition to that, it requires a trained expert to operate and interpret its results making it impossible for patients to directly use the device.

%% file: Chapter3.tex
\chapter{Device Setup and Experimental Analysis on Phantoms}
\label{chapter3}
This chapter\footnote{Work included in this chapter was done jointly in collaboration with Alejandro Velazquez Alcantar.} goes through the process of developing a NIRS based optical sensor and covers initial experiments done using a simple optical tissue phantom to validate the device. Further this chapter entails the process of selecting 970nm as the preferred wavelength for the optical probe used in this study and covers details on \textit{ex vivo} experiments using porcine bladder to test the device with a more realistic bladder environment. 

\section{Device Setup}
The developed device comprises of two parts- embedded hardware and an optical probe. Probe is placed over the outer skin in NIRS reflection mode (section \ref{modeofOperation}) to get information about the tissue structure beneath it. On the other hand, embedded hardware is used to control probe's optode system like driving the light emitting diode (LED) and recording the light after diffused reflectance as measured by detector photodiode (PD).

\subsection{Optical Probe}
\label{oldOpticalProbe}
As human tissue is a highly scattering medium (high signal attenuation), getting reasonable information from the deeper tissue such as location of bladder or amount of urine present inside the bladder is highly dependent on the selected light source. In this probe, a high-power light emitting diode (LED) with a peak wavelength of 970nm and a viewing angle of $\pm10^{\circ}$ is used. The experimental study leading to choice of 970nm as the preferred wavelength is covered later under section \ref{WavelengthSelection}. Narrow viewing angle helps to reduce the impact of photons residing in superficial layers of tissue structure. The probe is also covered with black tape to help reduce the number of escaped photons getting reflected back into the tissue structure and hence, further reducing the number of detected photons coming from superficial tissue layers.

As mentioned earlier, measuring a light signal that contains information about deeper tissues can be accomplished by observing the diffuse reflectance at a far distance from the originating light source. However, as source-to-detector (SD) distance increases, the amount of light seen at photodiode (PD) decays dramatically. Therefore, using photo-detectors with large active-areas help to capture as many photons as possible at these large SD separations. As the size of detector increases, its cost also increases. In this setup, to balance both cost and functionality a PD with an active area of 6.25mm$^{2}$, sensitive to 970nm wavelength and having a viewing angle of $\pm60^{\circ}$ was used. Contrary to emitters, detectors with large viewing angle is preferred so as to capture as many photons escaping the tissue as possible. 

As tissue layer between bladder mimic and optode in experiments covered under section \ref{phantomExperiments} is roughly 2cm, therefore, SD distance for this probe was picked to be 4cm following the principle - depth of photon penetration is approximately half the SD distance \citep{van1992experimentally, zonios2006modeling}.

\subsection{System Architecture}
\label{sec:sysarch}
\begin{figure}[h]
    \centering
    \includegraphics[width= 4in]{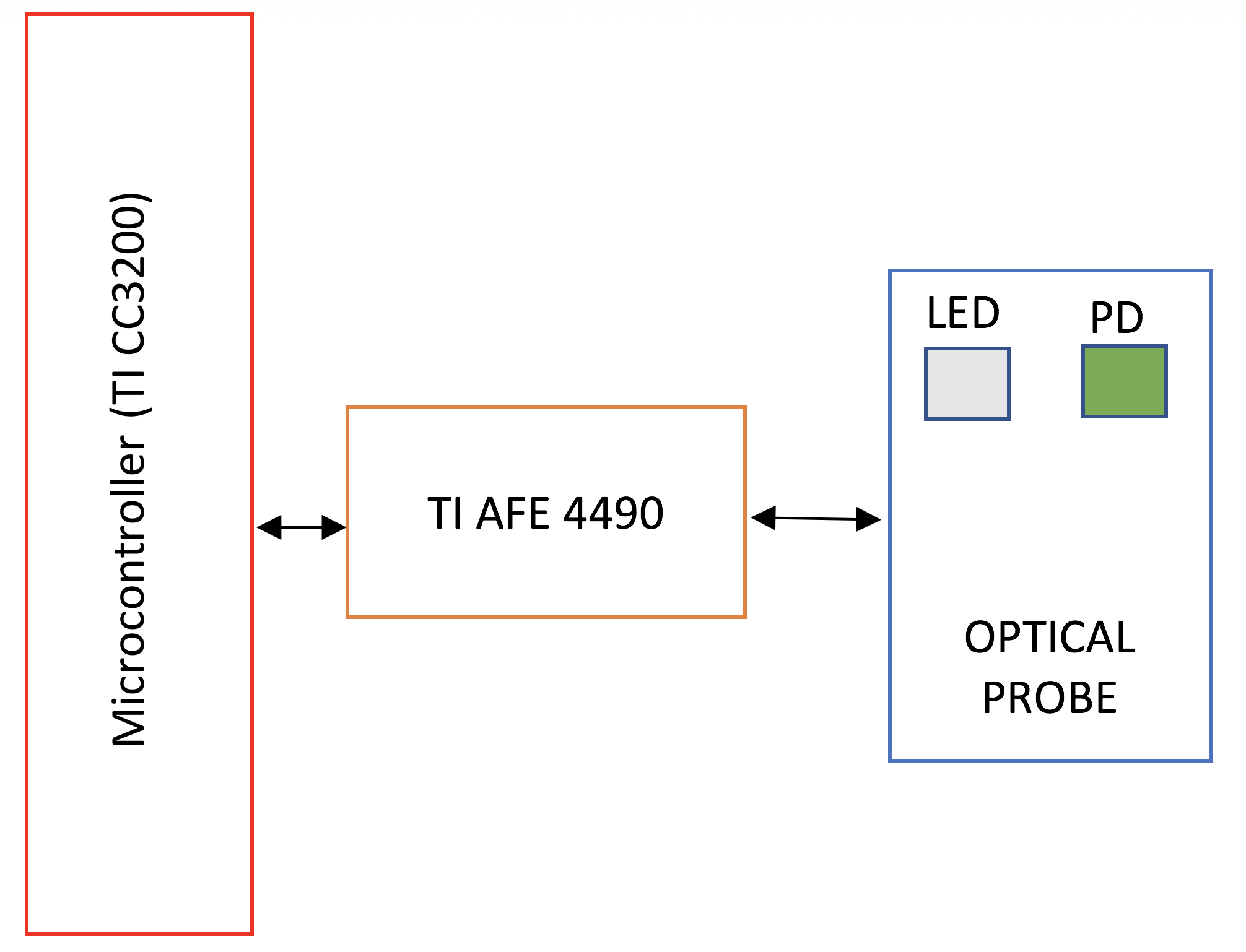}
    \caption{High-level system architecture}
    \label{fig:oldsysarch}
\end{figure}

The embedded device consists of Texas Instruments (TI) CC3200 microcontroller which is connected to a TI analog front-end chip called AFE4490. The microcontroller interacts with the optical probe via AFE chip. Figure \ref{fig:oldsysarch} gives high-level overview of system architecture. 

%LED Control
AFE is capable of controlling upto two LEDs independently with a programmable LED ON-time which gives the flexibility of operating LEDs in different modes like always-on, pulsating etc. Input current to LED is also programmable and has an 8-bit resolution for a fine control. In this setup, AFE is used to control only a single 970nm LED operated in pulsating mode. The pulsating mode is used to alternate between measuring ambient light present in the system and measuring light at detector PD when the emitter LED is turned on.

%Photodiode Control
Photodiode (PD) translates the detected photons exiting tissue into the electric current using photoelectric effect. Many electrical measurement components use voltage signals rather than current signals, to buffer, convey and manipulate information. To take advantage of this, current generated by photodiode is passed through a transimpedance amplifier (with programmable gain settings) to convert it into a voltage signal. To account for a variety of environmental settings, this signal then goes through a user-set ambient light cancellation system. As mentioned earlier, this setting can be estimated as taking measurement at detector when LED is turned off, and then fed back into the system to account for the level of ambient light seen. Output of this section is then buffered and fed into a 22-bit analog-to-digital converter (ADC) using a sample-and-hold circuit. 

Multiple ADC recordings are recorded and averaged over time to improve system accuracy. These measurements from ADC are carefully synchronized to the command and data handling subsystems to provide context to readings.

\section{Experiments}
\label{phantomExperiments}
\subsection{Study using Optical Tissue Phantom}
To test validity of above described system, a simple optical phantom was developed. Since 91-96\% of urine is composed of water \citep{rose2015characterization}, goal of this study was to detect the diffuse-reflected light coming from 970nm LED source (driven at 200mA) after it goes through the medium. Another important goal of the study was to test signal strength under various volumes of water. Change in signal strength with different volumes of water at PD would mean that the system is successfully able to distinguish between changes in water volume.  
%142mW of radiated power.    
%
\subsubsection{Setup}
\label{sec:opticaltissueSetup}
\begin{figure}[h]
\centering%
\begin{subfigure}[b]{0.5\linewidth}
    \centering
    \includegraphics[width=3in]{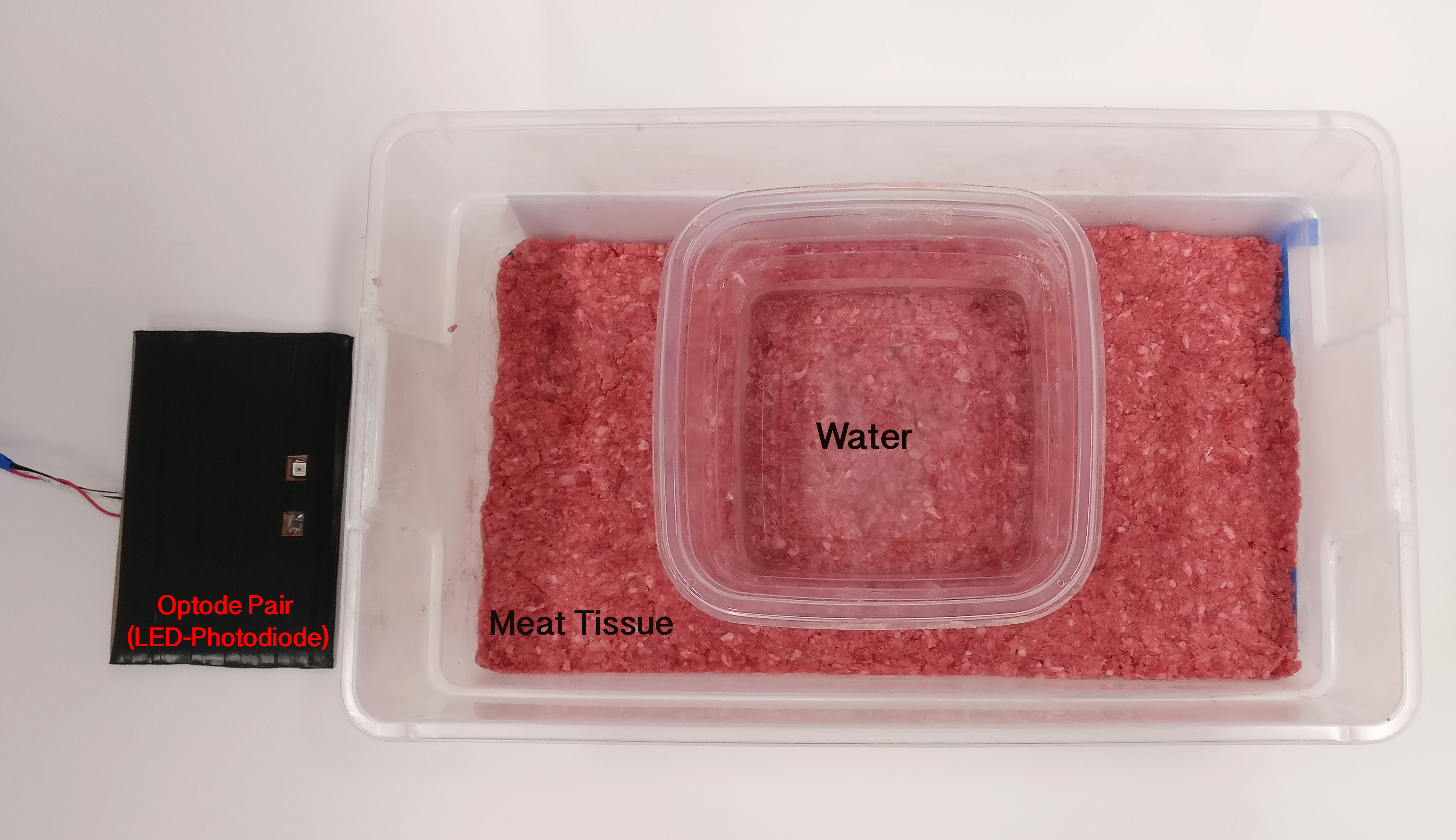}
    \caption{\label{fig:6.1}}
  \end{subfigure}%
  \begin{subfigure}[b]{0.5\linewidth}
    \centering
    \includegraphics[width=3in]{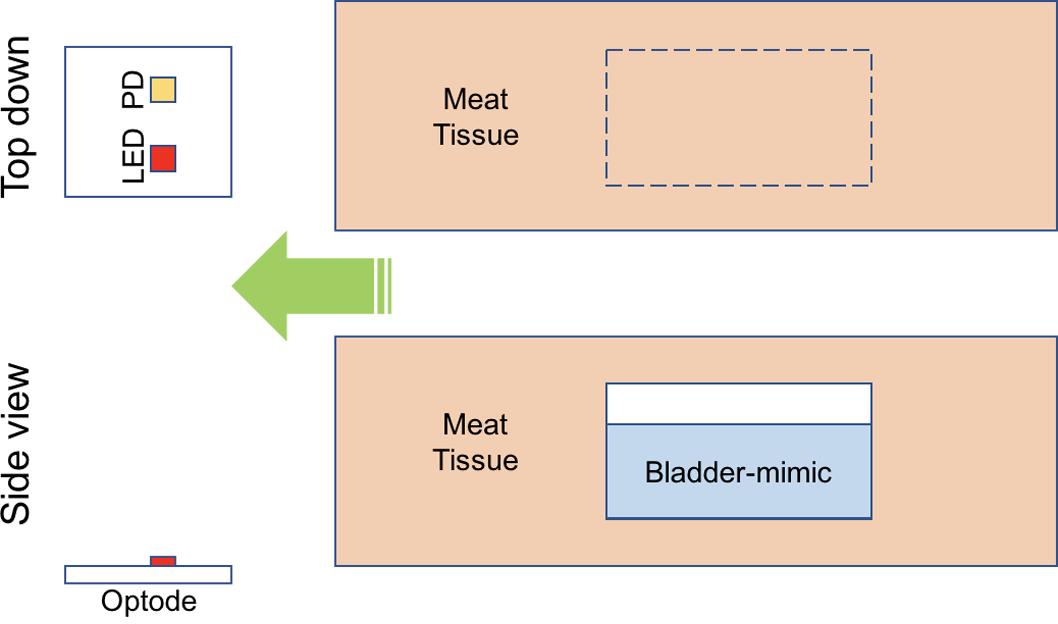}
    \caption{\label{fig:6.2}}
  \end{subfigure}%
  \caption[Experimental setup of Optical Tissue Phantom]{(\subref{fig:6.1}) A picture showing container of water representing the bladder before being covered with food grade bovine muscle and fat representing the surrounding tissue \citep{dfong2018lepsbv}. (\subref{fig:6.2}) Phantom moves across optode pair taking readings every cm across the length of phantom \citep{dfong2018lepsbv}.}
\label{figure6}
\end{figure}

The optical phantom was developed to mimic the gross anatomy, physiology and optical properties of human bladder along with its surrounding tissue as shown in figure \ref{figure6}. Bladder is represented with an 8*8cm container filled with water whereas length of entire phantom is 22cm. A 2cm thick layer of tissue consisting of bovine muscle and fat was present between the optical probe and bladder mimic. After filling the container with water, surrounding area to the bladder mimic was also covered with the same tissue. 

Phantom was moved across the stationary optode pair taking reading every centimeter. Since the normal capacity of human bladder is between 400-500ml, measurements were taken at water volumes- 100ml, 300ml and 500ml.  

\subsubsection{Results}
\begin{figure}[h]
    \centering
    \includegraphics[width=5in]{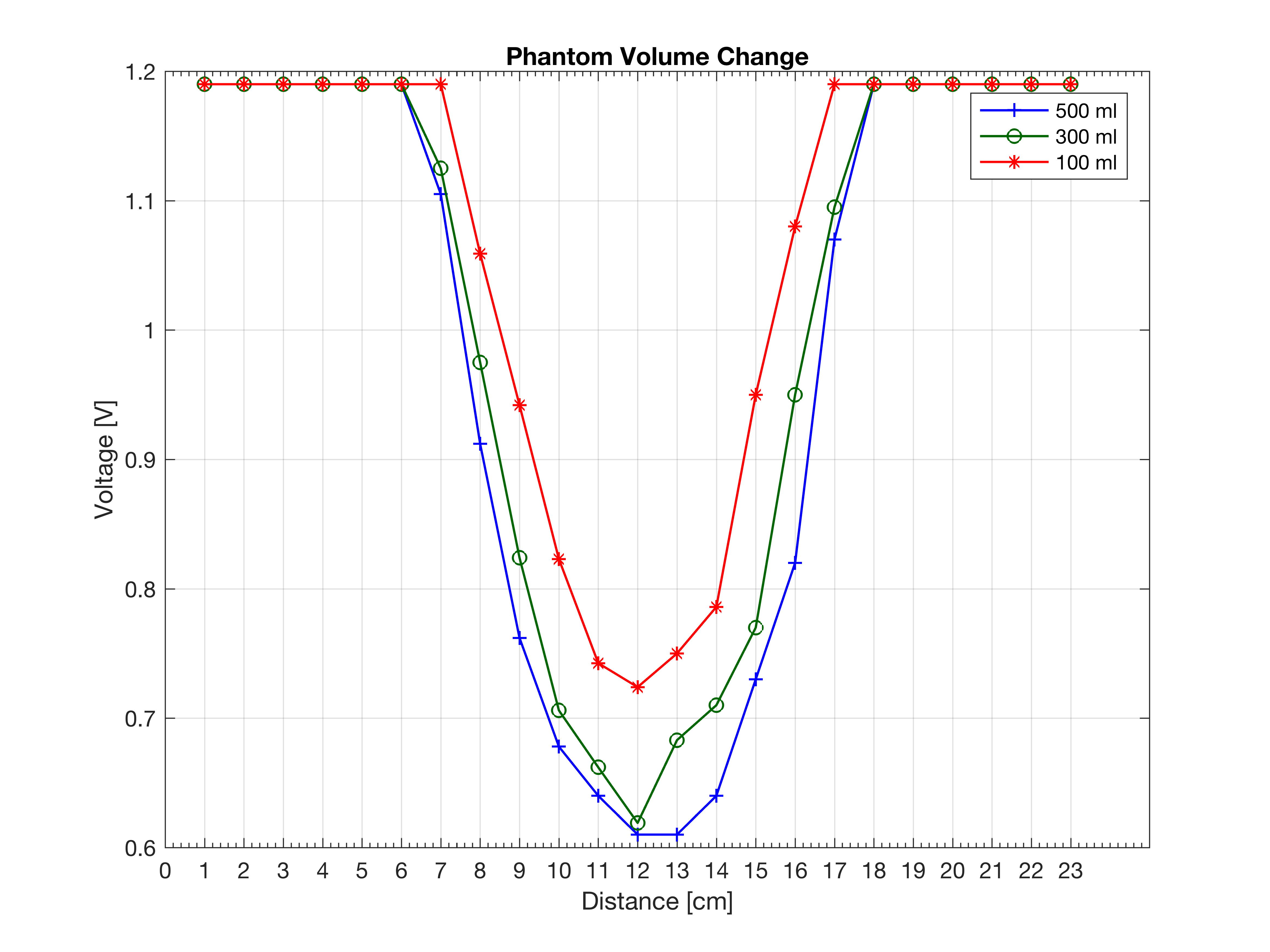}
    \caption[Results of Optical Phantom]{Measurements performed on the optical phantom over three volumes of liquid (100ml, 300ml, 500ml) using a 970nm LED. As expected, as amount of water in the bladder increases, light intensity drops \citep{dfong2018lepsbv}.}
    \label{fig:OpticalPhantomGraph}
\end{figure}
Figure \ref{fig:OpticalPhantomGraph} shows results from the experiment described in section \ref{sec:opticaltissueSetup} for three levels of water volume. It was observed that as amount of water inside the bladder-mimic increases, there is a noticeable decrease in intensity of light at the detector PD. Intensity of light is represented in term of voltage [V] (described in section \ref{sec:sysarch}) on y-axis while x-axis shows location of probe as phantom moves across it. As total length of phantom is 22cm, readings shown in figure \ref{fig:OpticalPhantomGraph} at 0cm and 23cm depict the case when phantom is not covering optodes or in other words when medium for photons is air.  
\subsection{Wavelength Selection}
\label{WavelengthSelection}
As mentioned earlier, selecting the right wavelength is critical for NIRS applications to attain a reasonable optical signal and get  appropriate level of sensitivity. To test this, three LEDs having peak wavelengths at 890nm, 970nm and 1450nm respectively were tested, each of which has a peak for water absorption. These wavelengths were chosen to investigate the effect that different absorption coefficients have on the overall light intensity measured, in addition to the coefficient’s stability over small variations in wavelength. Investigating these variations in wavelength helps to reduce errors caused by slight shifts in the peak wavelength of the LED. Absorption coefficient for water at 890nm is 0.058cm$^{-1}$, 970nm is 0.481cm$^{-1}$, and 1450nm is 32.778cm$^{-1}$ \citep{kou1993refractive}. 

Each experiment conducted with different wavelength of LED used the same setup as described in section \ref{sec:opticaltissueSetup} with only difference being the amount of water inside the container being fixed at 300ml.
\subsubsection{Results}
\begin{figure}[h]
    \centering
    \includegraphics[width=5in]{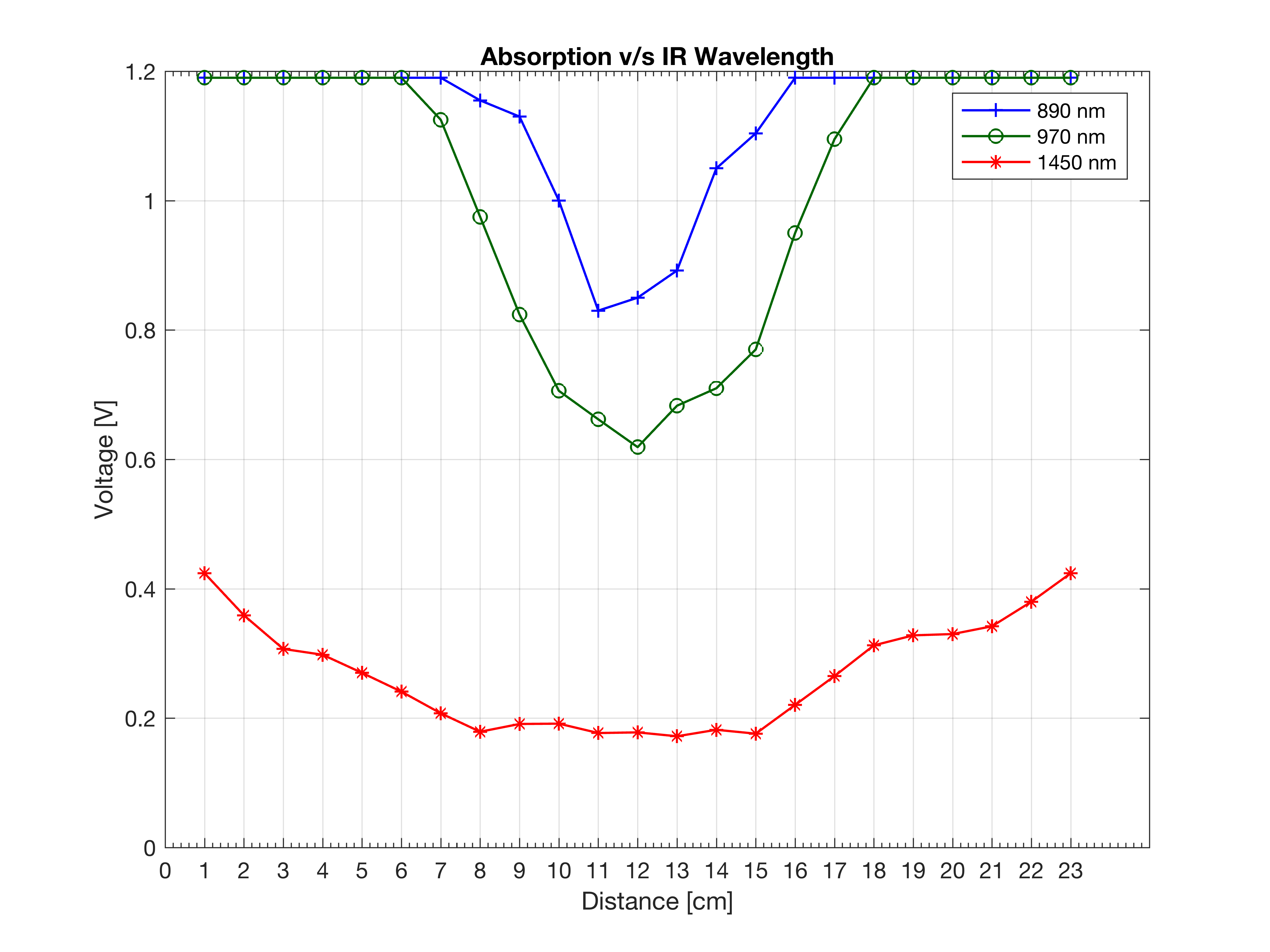}
    \caption[Measurement for wavelength selection]{Measurements performed on optical phantom over three wavelengths (890nm, 970nm, 1450nm). As expected, the depth of the light intensity signal (in Volts) follows the absorption coefficient for water at these wavelengths \citep{dfong2018lepsbv}.}
    \label{fig:WavelenghtSelection}
\end{figure}
Figure \ref{fig:WavelenghtSelection} shows light intensity (in volts) detected at PD for various wavelengths. It was observed that 1450nm wavelength is too sensitive to water degrading its ability to reach deeper tissue levels and hence, unable to predict the exact location of water in phantom. 970nm has optimal sensitivity for this setup giving a maximum resolution of 0.6V.
\subsection{\textit{Ex vivo} study using Porcine Bladder}
Findings from initial Optical Tissue Phantom experiment validates the idea, however in order to investigate the results in a more realistic bladder environment, \textit{ex vivo} experiments were done using porcine bladder. 

Choice of porcine bladder was made because of its anatomical and physiological similarities with the human bladder. In addition to that, it has a capacity of approximately 500-550ml which is comparable to human bladder. For these reasons, it is a popular experimental model for conducting any urology related research \citep{chen2015design}. Porcine bladder and intestines used in this study were obtained from the UC Davis Meat Lab.
\subsubsection{Setup}
\begin{figure}[h]
\centering%
\begin{subfigure}[b]{0.5\linewidth}
    \centering
    \includegraphics[width=3in]{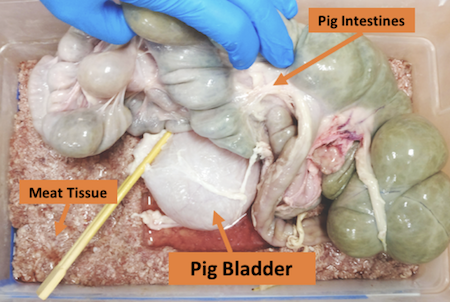}
    \caption{\label{fig:7.1}}
  \end{subfigure}%
  \begin{subfigure}[b]{0.5\linewidth}
    \centering
    \includegraphics[width=3in]{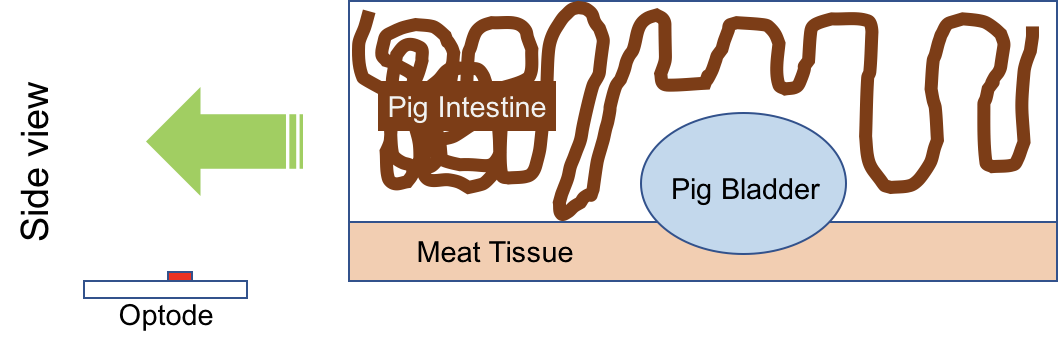}
    \caption{\label{fig:7.2}}
  \end{subfigure}%
  \caption[Setup of \textit{ex vivo} experiments using Porcine Bladder ]{\subref{fig:7.1}) A picture of \textit{ex vivo} setup showing the use of porcine bladder, large intestines to create a more realistic tissue model \citep{dfong2018lepsbv}. \subref{fig:7.2}) Phantom moves across optode pair taking readings every cm across the length of phantom similar to the Optical Tissue Phantom \citep{dfong2018lepsbv}.}
\label{figure7}
\end{figure}

The \textit{ex vivo} experiment was aimed to realistically mimic anatomy, physiology and optical properties of human bladder. Porcine bladder was placed in the centre of phantom. A 2cm layer of bovine muscle and fat was present at bottom of phantom between bladder and optode. Porcine bladder was surrounded by intestines in order to create a more anatomically accurate depiction of bladder environment. Bladder was filled with 200ml of water using a system of syringes, tubes, and clamps. A 970nm LED driven at 670mA was used to perform the measurements. Figure \ref{figure7} shows a detailed visual presentation of the setup. Similar to section \ref{sec:opticaltissueSetup}, phantom was moved across the stationary optode pair taking reading every centimeter.

\subsubsection{Results}
\begin{figure}[h]
    \centering
    \includegraphics[width=5in]{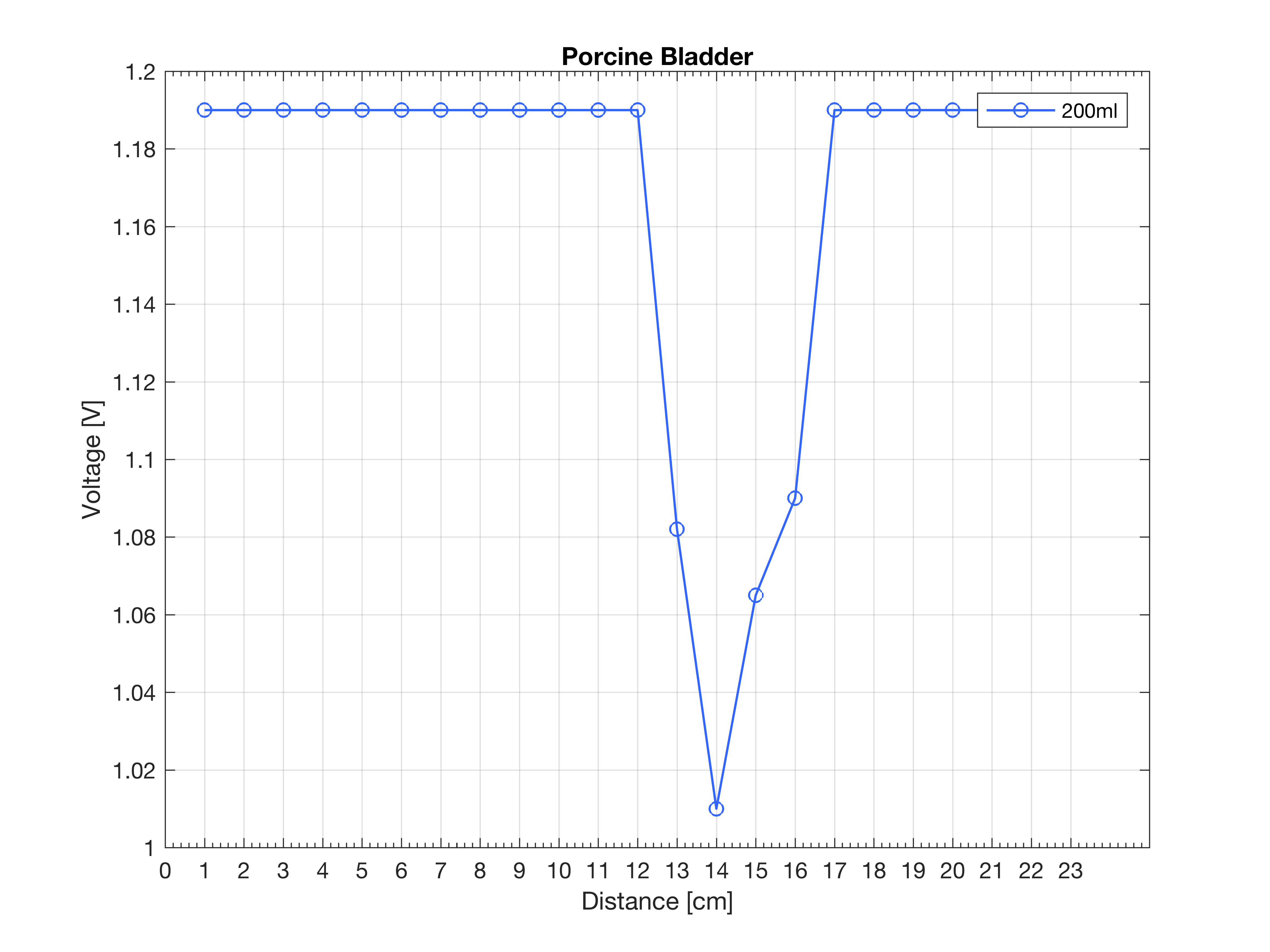}
    \caption[Results of \textit{ex vivo} experiment using Porcine Bladder]{Measurements performed on porcine bladder (filled to 200ml of water) using a 970nm LED. As the NIR light field passes over the bladder, characteristic drop in light intensity appears \citep{dfong2018lepsbv}.}
    \label{fig:PorcinePhantomGraph}
\end{figure}

Figure \ref{fig:PorcinePhantomGraph} shows a noticeable drop in light intensity at the location where porcine bladder is present inside the phantom. Intestines contain partially digested food, pancreatic juices, fecal matter etc. and despite that, NIRS based methodology using IR LED at 970nm wavelength is successfully able to detect water inside bladder travelling through a 2cm thick layer of human tissue. %TODO: Make this better

%% file: Chapter4.tex
\chapter{Experimental Analysis on Human Subjects}
\label{chapter4}
Human body is full of variations, having different body type, shape and size. The experimental analysis covered under section \ref{phantomExperiments} proves that an NIRS based methodology using 970nm wavelength is successfully able to detect water through a 2cm thick layer of tissue in a controlled environment. Making a transition from experimental setups to testing on human subjects commanded major improvements in the system architecture so that the system could be robust enough to handle the variations in the physical characteristics of bladder like shape, size, capacity, location, dynamics etc. 

\section{Updated Probe Design}
\label{currentprobedesign}
In order to develop the updated device, following considerations were taken into account-

\begin{itemize}
    \item \textbf{Wearable System}\\
    For testing on human subjects, the device had to be flexible and light-weight so that it could be easily wrapped around the abdominal area. For NIRS applications, light source and detector should be in flushed contact with skin to prevent any light loss and get reasonable information but it may also introduce thermal effects between the tissue and probe. Flexibility of the probe thus would enable it to adjust to any body shape.
    \item \textbf{Bladder Shape, Size and Location}\\
    Location, shape and size of human bladder varies slightly from person to person. The probe described earlier in section \ref{phantomExperiments} had only one LED-PD pair and hence the chance of missing the bladder if the probe is not placed carefully was extremely high. Another important aspect to account for is the variable waist sizes resulting in variation of tissue thickness (present between the probe and the bladder). Hence, developing a system with multiple LED-PD pairs has the ability to cover a large abdominal area and to measure the diffused reflectance of input light for variable penetration depths.
    \item \textbf{Energy Consumption} \\
    As human body is a highly scattering and dynamic medium, driving the LEDs at a higher power would increase the probability of photons being detected at PD. However, as the input power increases, heat generated by the system also increases which may lead to arbitrary behavior of embedded systems. Increasing the number of LEDs and PDs on probe also increases the energy budget required for switching in-between them. A steady power source is also desirable so as to increase reliability of the data collected by the system.
    \item \textbf{Flexible Design}\\
    A flexible design that is scalable and adapts as the user requirements change is always desirable. For example, having fine control over data acquisition, programmable gain settings, ability to switch mode of operation etc.
\end{itemize}

\subsection{Optical Probe}
 Taking the above mentioned design considerations into account, system was upgraded\footnote{Work included in this section was done jointly in collaboration with Daniel Fong.} by adding eight 970nm LED-PD pairs to the probe made with copper-coated polyimide (Kapton, DuPont) to ensure full coverage of bladder. Polyimide, a flexible bio-material that is used in various flexible electronics \citep{chen2015breathable}, was used instead of using a conventional rigid substrate. It can be etched or milled to create electrical traces, thereby allowing the integration of LEDs and PDs with the substrate. Any exposed copper was then covered with polyimide tape to prevent over-time corrosion. 
 
\begin{figure}[h]
    \centering
    \includegraphics[width= 5in]{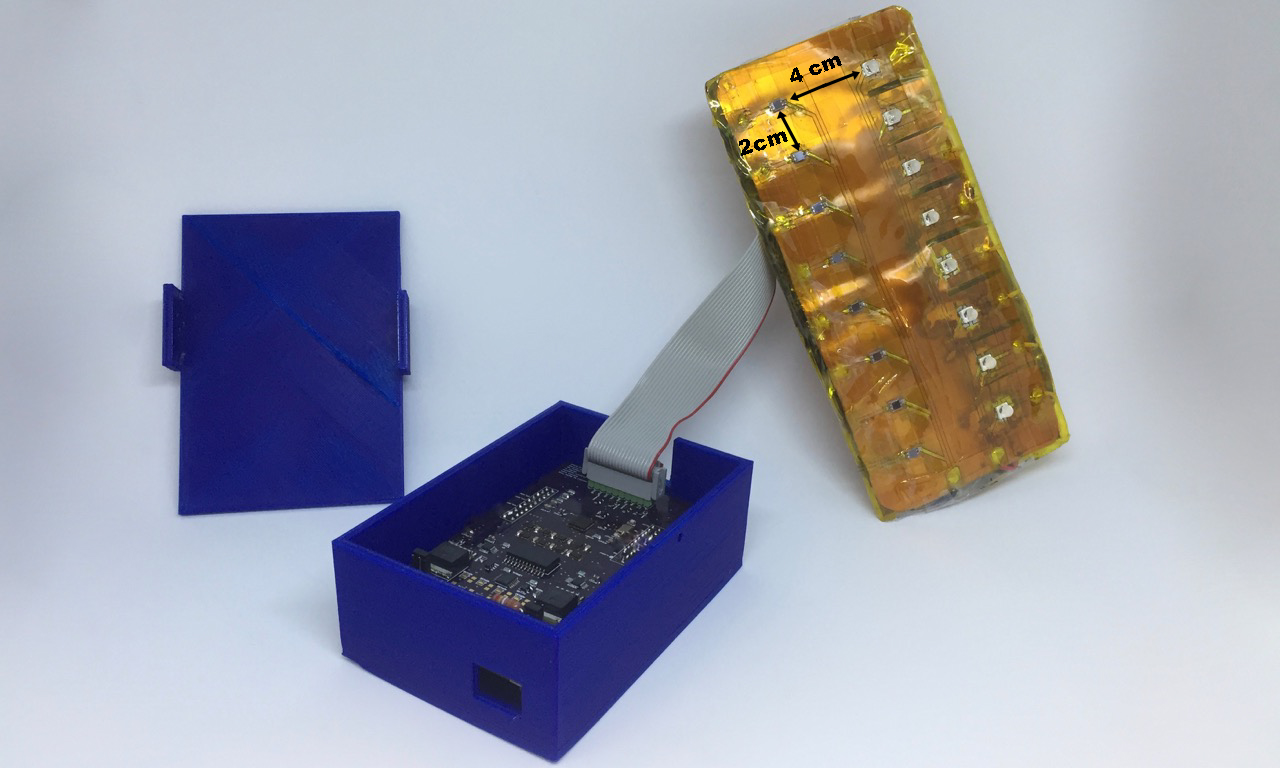}
    \caption[Real life picture of the device]{A real life picture showing both parts of the setup - a wearable, flexible, non-invasive optical probe; optode control system that actuates between the light emitters and records the resulting diffuse reluctance placed in a 3D printed box \citep{df2018lepsbv2}.}
    \label{fig:currentdevice}
\end{figure}

 Since human tissue is a highly-scattering medium, when a photon enters tissue it is quickly scattered from its initial trajectory. This means that more photons will likely reside close to its entry point, and are more probable to exit the tissue closer to the source. However, photons that exit the tissue further from the emitter are more likely to have traversed deeper into the tissue. Keeping in mind that the total number of photons that exit at that distance is much lower, this setup uses a SD distance (distance between LED and PD) of 4cm to keep things consistent with the probe used for optical phantoms. Each optode pair is then spaced 2cm away from each consecutive pair as shown in Figure \ref{fig:currentdevice}. Taking into account the international standard of safety limit for operation on human subjects \citep{din60601,international2006iec} the LEDs were driven at 800mA to achieve the best possible photon penetration. 
 
\subsection{System Architecture}
\label{sec:newsysarch}
 The probe control system can be distributed into 2 main modules- driving LEDs and recording diffused reflectance as seen at PDs. To control the updated probe, significant changes had to be made on the device side for enabling new features like LED-PD actuation and synchronization, data acquisition etc.
 
  \begin{figure}[h]
    \centering
    \includegraphics[width= 5in]{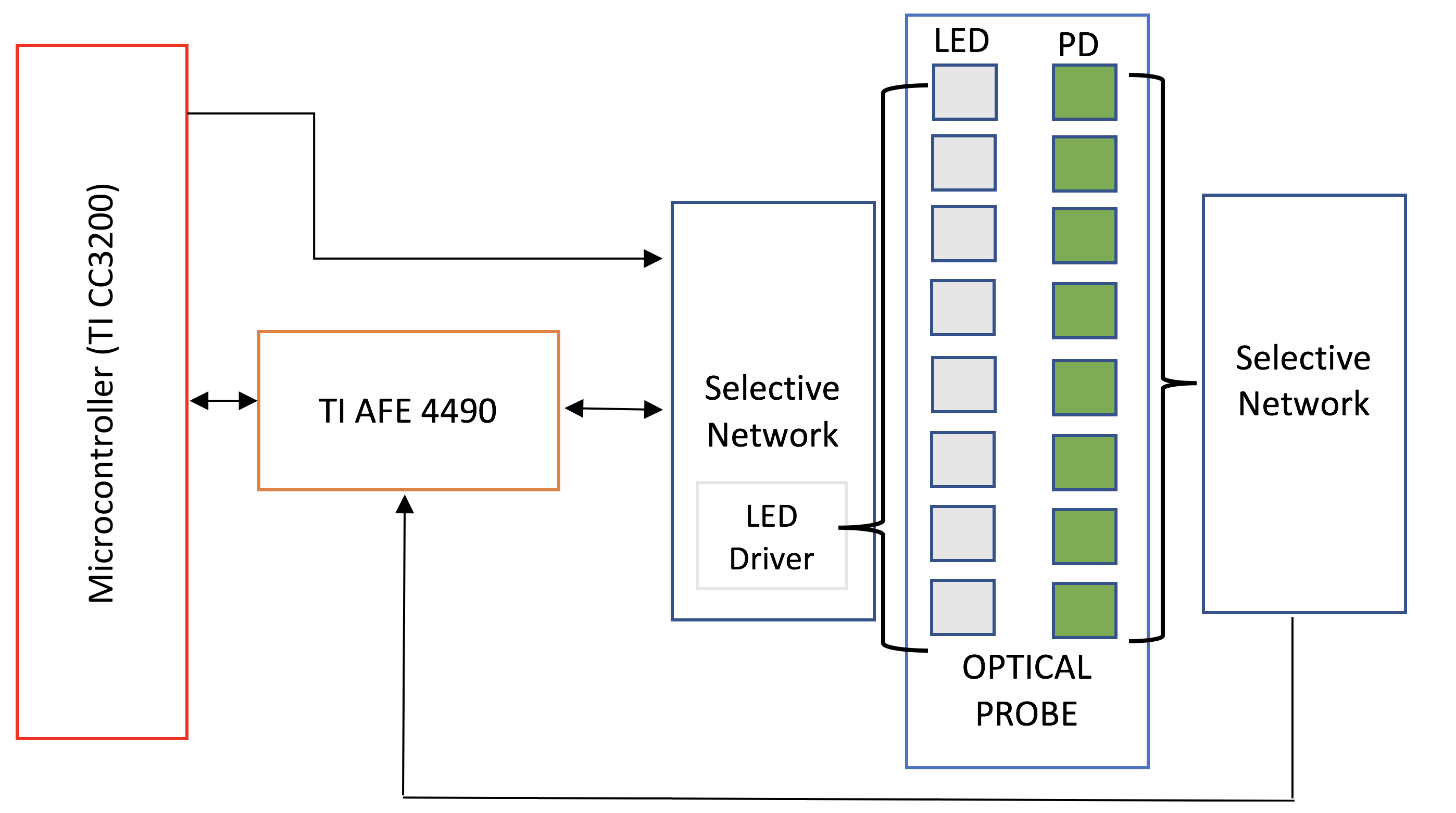}
    \caption[Updated System Architecture]{High-level system architecture for current probe having multiple LEDs, Photodiodes.}
    \label{fig:newsysarch}
\end{figure}
 
As mentioned earlier, its desirable to operate LEDs at high-power for deep-tissue measurements but it often requires LED drivers that can handle large amounts of current. In order to develop a flexible system that could tune as per user requirements, multiple LEDs with programmable current settings and high power LED drivers had to be used. These drivers take up a lot of space, in addition to that, the TI AFE chip could only support a maximum of two LEDs as mentioned in section \ref{sec:sysarch}. To overcome these limitations, one high power LED driver was used in conjunction with series of high-current capable switches to control actuation between multiple LEDs. This removed the requirement of having multiple switches but allowed only 1 LED to be actuated at a given time. Current driver uses a set value of resistor to control the amount of current fed into the LED which is achieved by using a digital potentiometer. It is programmed using a microcontroller which receives user settings and interprets them in real-time. Similar to LED actuation, PD switching is achieved by a series of low-resistance switches between PDs and transimpedance amplifier, to select the detector of interest and feed its signal into a single set of aforementioned detection components. Figure \ref{fig:newsysarch} shows a high-level overview of the updated system architecture.
 
Microcontroller through a single channel is thus used to activate the desired LED and PD pair. Collected data after passing through the programmable ADC is off-loaded to a laptop via a USB connection which is also used for collecting user initiated requests and powering the device. The off-loaded data is further analyzed to discover underlying patterns. 

\begin{table}[!ht]
\centering
\caption{Comparison of the developed system with that of \cite{molavi2014noninvasive} }
\label{tab:movali}
\begin{tabular}{|c|c|c|}
\hline
& \textbf{\cite{molavi2014noninvasive}} & \textbf{Our System}  \\
 \hline
 \textbf{No. of Optode Pairs}  & 1 & 8 \\
 \hline
 \textbf{Emitter Wavelength}  & 950nm & 970nm    \\
 \hline
 \textbf{Emitter Operation}  & 370mA, 60mW & 800mA, 586mW  \\
 \hline
 \textbf{Detector Active Area}  & 5.22mm$^{2}$ & 6.25mm$^{2}$   \\
 \hline
 \textbf{Detector Responsivity}  & 0.45A/W & 0.56A/W  \\
 \hline
 \textbf{SD Gap}  & 3cm & 4cm  \\
 \hline
 \textbf{AFE Gain}  & 6x10$^6$V/A &  $\approx$ 1x10$^4$V/A\\ 
 \hline
\end{tabular}
\end{table}

%comparison with molavi. 
Since, the technique of using NIRS to sense change in bladder volume used in this study is closest to that of \cite{molavi2014noninvasive}, table \ref{tab:movali} shows the specification comparison between the systems used in each study. Higher SD gap enables the developed system to penetrate deeper tissue layer with an emitter wavelength of 970nm which is more sensitive to water as compared to 950nm. Higher operating current and optical power generated at the emitter translates to more photons reaching the detector. Higher detector responsivity points to higher current produced by the PD per watt of optical power it receives.     

\section{Volunteer Enrollment and Data Collection}
\label{IRBdetails}

\begin{figure}[h]
\centering%
\begin{subfigure}[b]{0.5\linewidth}
    \centering
    \includegraphics[width=3in,height=5in]{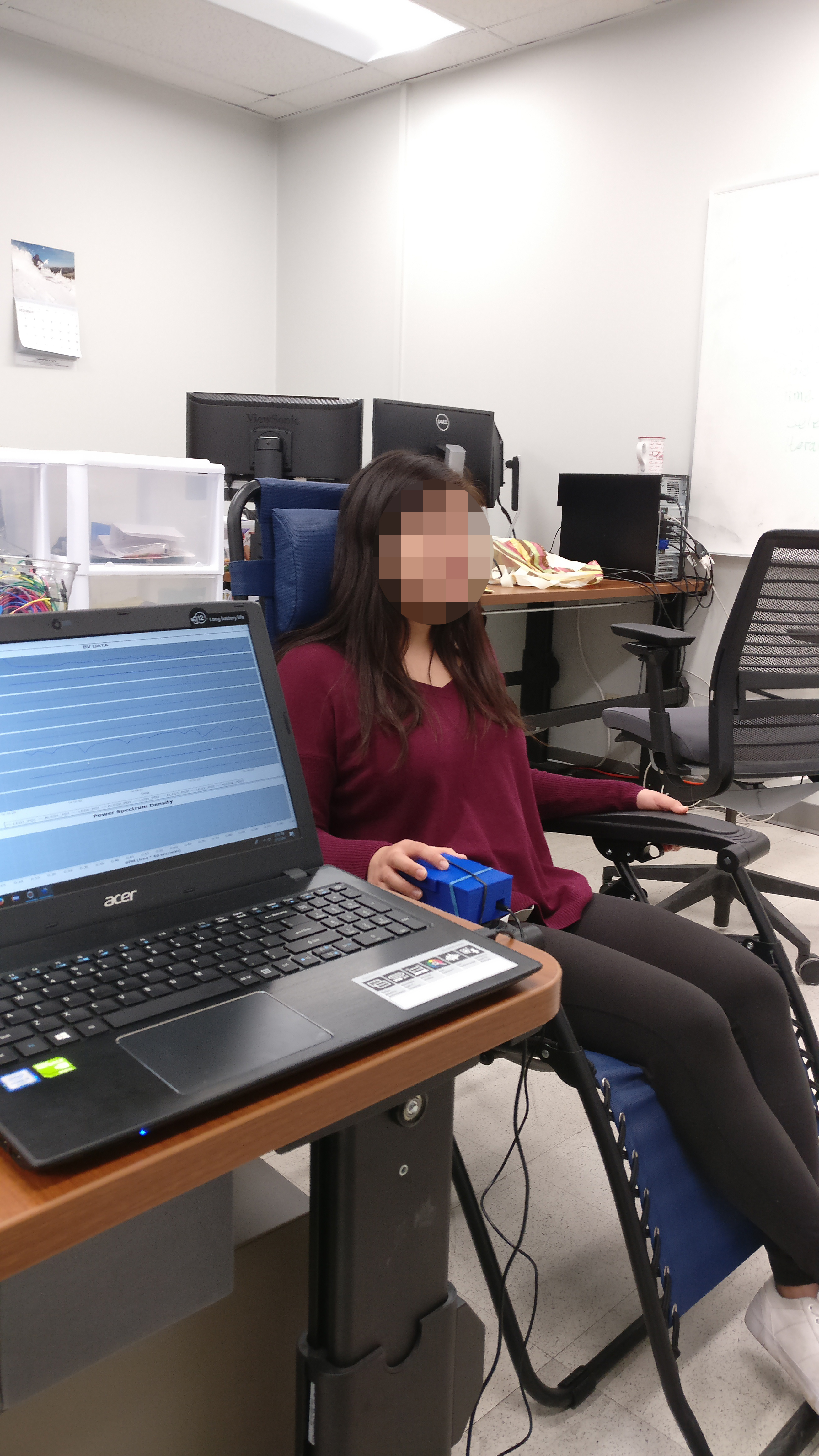}
    \caption{\label{fig:humantrial}}
  \end{subfigure}%
  \begin{subfigure}[b]{0.5\linewidth}
    \centering
    \includegraphics[width=3in,height=5in]{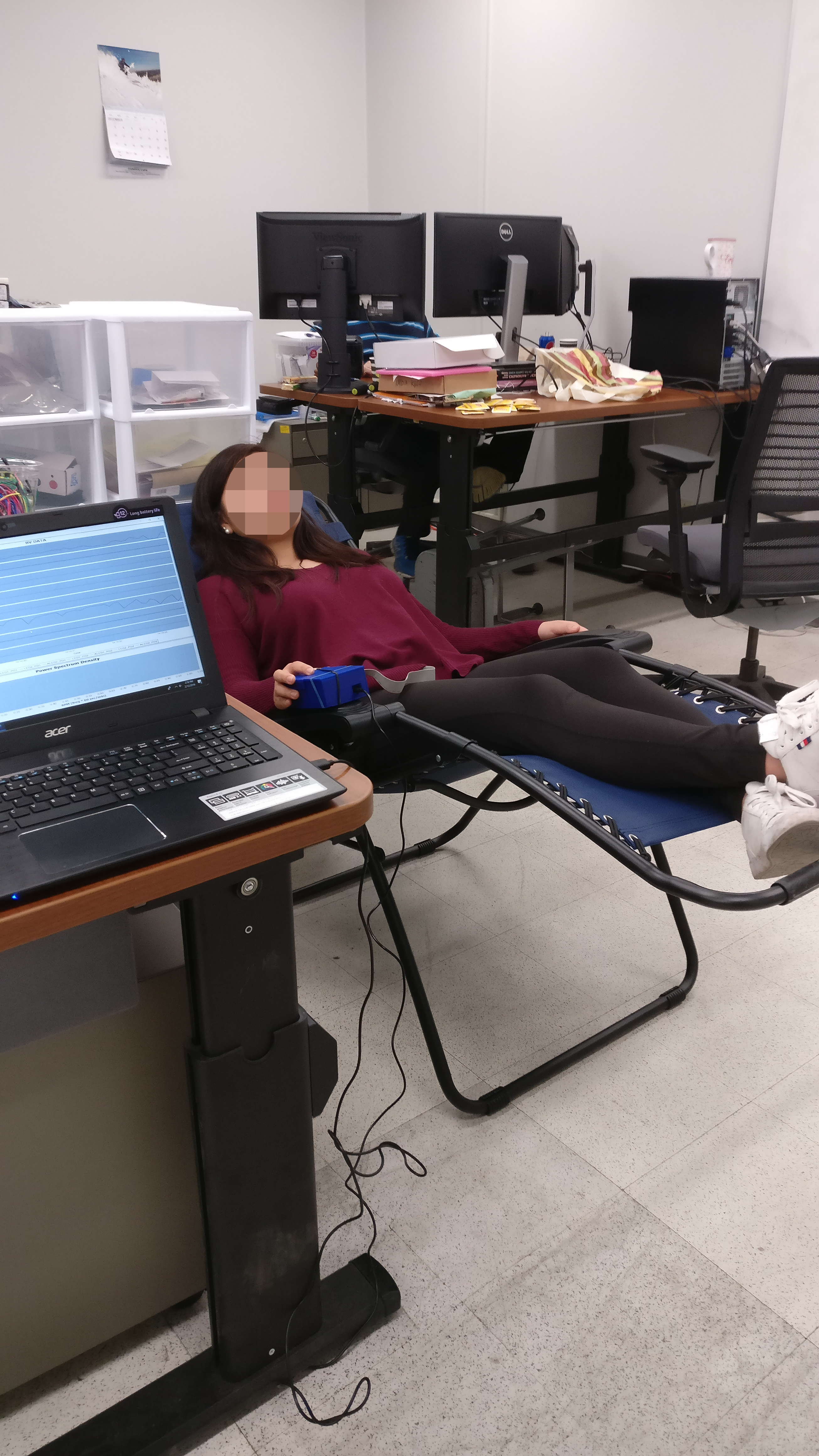}
    \caption{\label{fig:humantrial1}}
  \end{subfigure}%
  \caption[Volunteer undergoing bladder volume measurement study]{ Volunteer undergoing bladder volume measurement trial using the device described in section \ref{currentprobedesign}. \subref{fig:humantrial}) Readings being collected with volunteer in sitting position. \subref{fig:humantrial1}) Readings being collected with volunteer in reclined position.}
\label{figure10}
\end{figure}

Adult healthy subjects were enrolled in the study after it was approved by UC Davis Institutional Review Board (IRB). Subjects were contacted through emails and flyers posted across the UC Davis campus. For each subject volunteering for the study, probe was placed on the outer skin, (more details on probe placement are covered in section \ref{sec:probeplacement}) and readings were taken in 3 positions - \textit{standing}, \textit{sitting} and \textit{reclined} for 1 minute each. Readings collected over this time were later averaged to reduce the error due to movements or other external factors. Idea behind different positions was to cover some basic scenarios that a person suffering from Neurogenic Bladder Dysfunction may go though in day-to-day life. To measure the amount of urine present inside bladder, subjects were asked to void in a urine specimen collection container. State of bladder just before voiding is referred to as \textit{full-bladder state} and it was assumed that being healthy subject the volunteer would void their bladder to full and thus, volume of liquid inside their bladder post-void was assumed to be zero ml or \textit{empty-bladder state}. For each subject other biographical information like age, gender, height, weight and waist (from hip to hip) was also recorded. Figure \ref{figure10} shows a volunteer undergoing bladder volume measurement study.

As human body is an extremely variable system, biographical information thus collected will prove useful for the future goals of this study where using machine learning algorithms, a model can be trained for tuning the device specifically to meet requirements of each individual subject (more details about this is covered in section \ref{futurework}).
\subsection{Probe Placement}
\label{sec:probeplacement}
\begin{figure}[h]
    \centering
    \includegraphics[width= 5in]{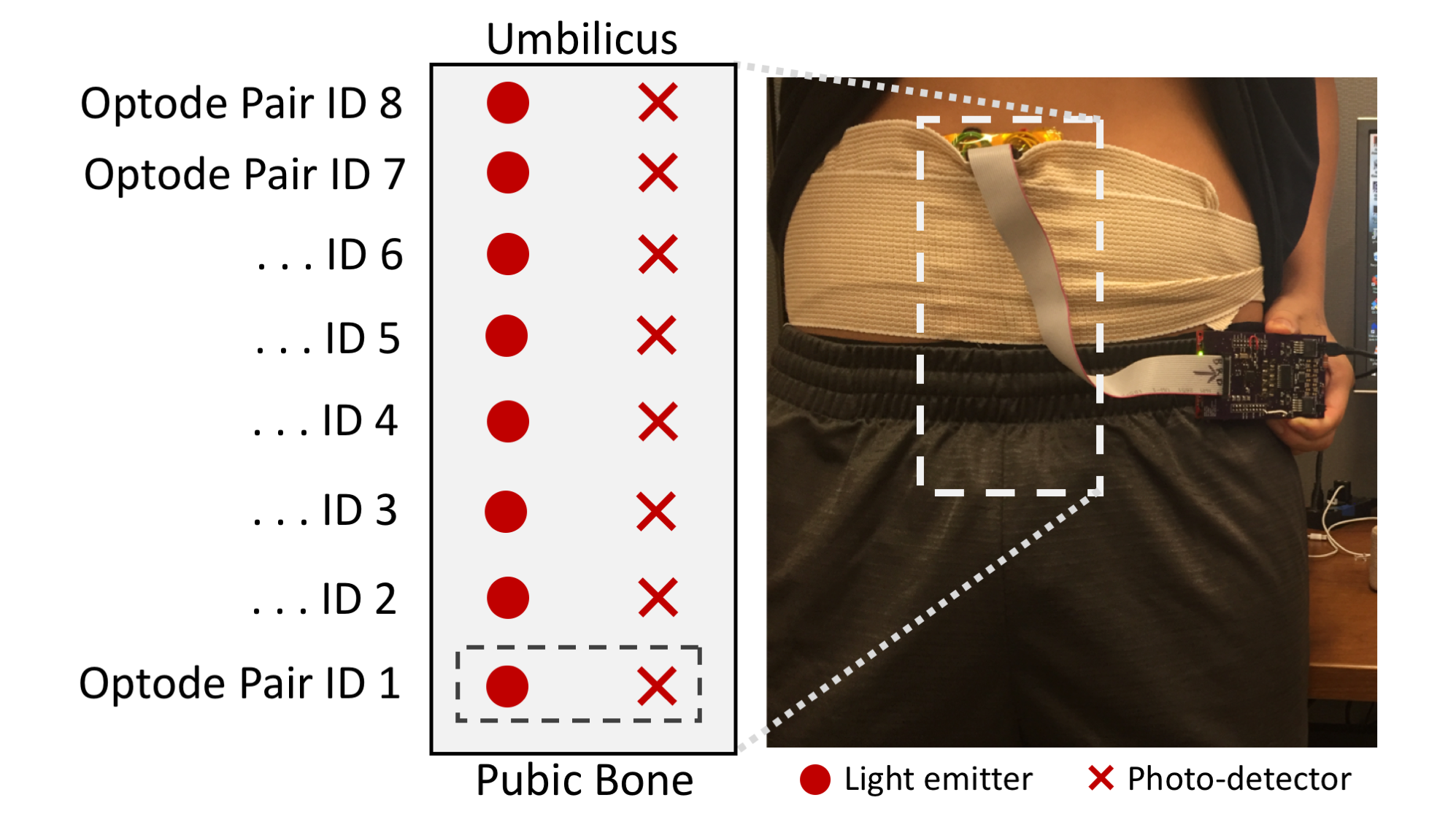}
    \caption[Probe placement on the participants]{Picture outlining the normal placement of probe on a participant\citep{df2018lepsbv2}.}
    \label{fig:sensorplacement}
\end{figure}
For collecting data on human subjects probe was placed on lower abdomen between umbilicus and pubic bone as shown in Figure \ref{fig:sensorplacement}. Optode pairs are numbered from 1 to 8, ID number 1 being closest to the pubic bone whereas ID number 8 is closest to the umbilicus. 

As mentioned earlier, contact between optode and skin is very important for NIRS measurements to avoid any signal loss. To maintain a flushed contact with the skin, probe was placed under all clothes and a self-adherent wrap was used to apply pressure to keep the sensor at a fixed location. The wrap made sure that sensor remained stationary as the subject moved around for measurements.
\section{Experiments}
\label{humanexp}
This section covers details on the experiments that were conducted on healthy subjects after approval by the IRB. Goal of these experiments was to evaluate feasibility of the developed device on human subjects. 

For experiments included in this section, signal detected at PD due to emitted LEDs is referred to as \textit{desired signal} while detected signal as a result of any other light source, e.g ambient light is referred as \textit{noise}. \textit{Desired signal} can be separated from the detected signal by subtracting ambient voltage value.

\subsection{Full and Empty bladder state comparison with multiple volunteers}
\label{multivolunteer}

10 adult healthy volunteers were selected to test the developed device. To maintain volunteer confidentiality each volunteer was given a code namely, \textit{UCD101} to \textit{UCD110} depending on the order in which they signed-up.

Each volunteer came in for the experiment with a \textit{full-bladder} and then probe was placed on the abdomen taking readings in aforementioned positions as explained in section \ref{IRBdetails}. After this, the subject was asked to void in a urine specimen collection container, to quantify the amount of urine present inside the volunteers bladder. Post-voiding the same set of readings were repeated for an \textit{empty-bladder state}. In order to minimize error caused due to sensor movement, probe remained attached to the volunteer throughout the experiment and was only removed after both pre- and post-void data was successfully recorded. Measurements were recorded at each of the optode pairs present on the probe along with ambient light readings, to achieve an overall sampling rate of 1Hz. As physiological change in bladder volume is a slow process, this sampling rate is appropriate for detecting those changes.

Goal of this experiment was to test if intensity of light at the detector increases as volume of liquid inside the bladder decreases from full- to empty-bladder state, as seen with phantom experiments.
\subsubsection{Results}
\begin{figure}[H]
\centering%
\begin{subfigure}[a]{0.49\linewidth}
    \centering
    \includegraphics[width=\linewidth]{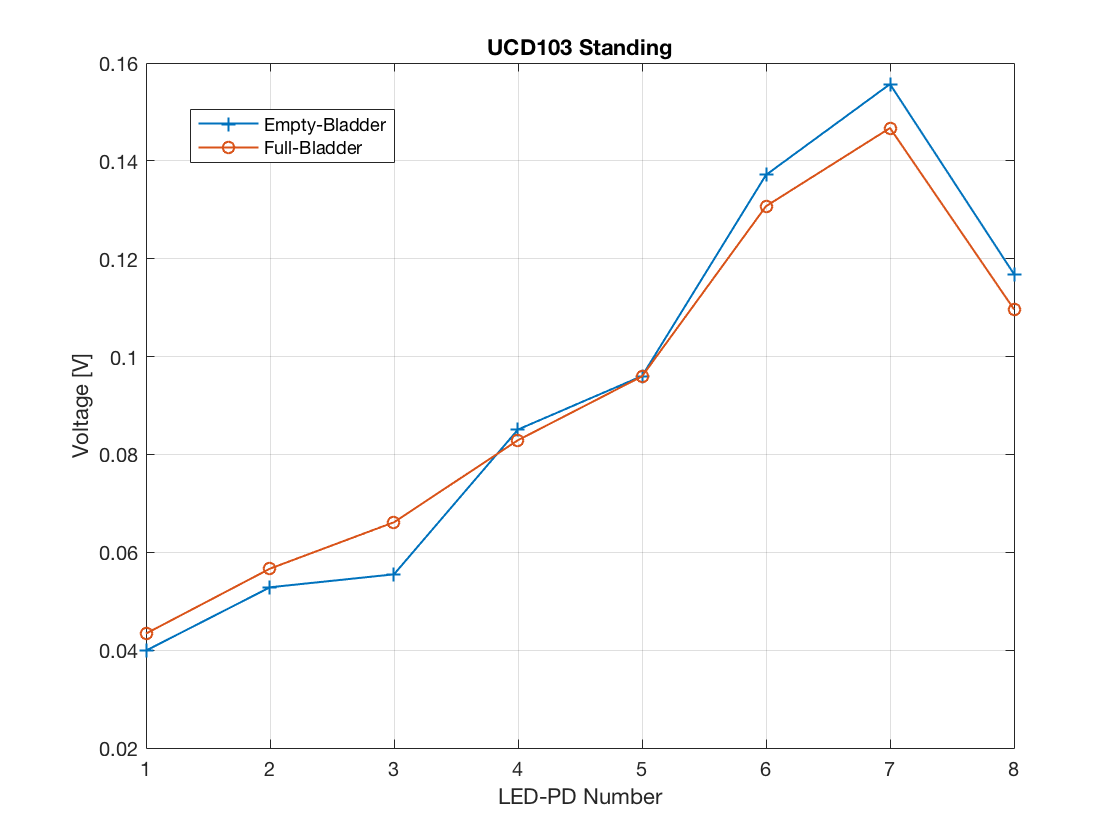}
    \caption{\label{fig:mutli_standing103}}
  \end{subfigure}
  \begin{subfigure}[a]{0.49\linewidth}
    \centering
    \includegraphics[width=\linewidth]{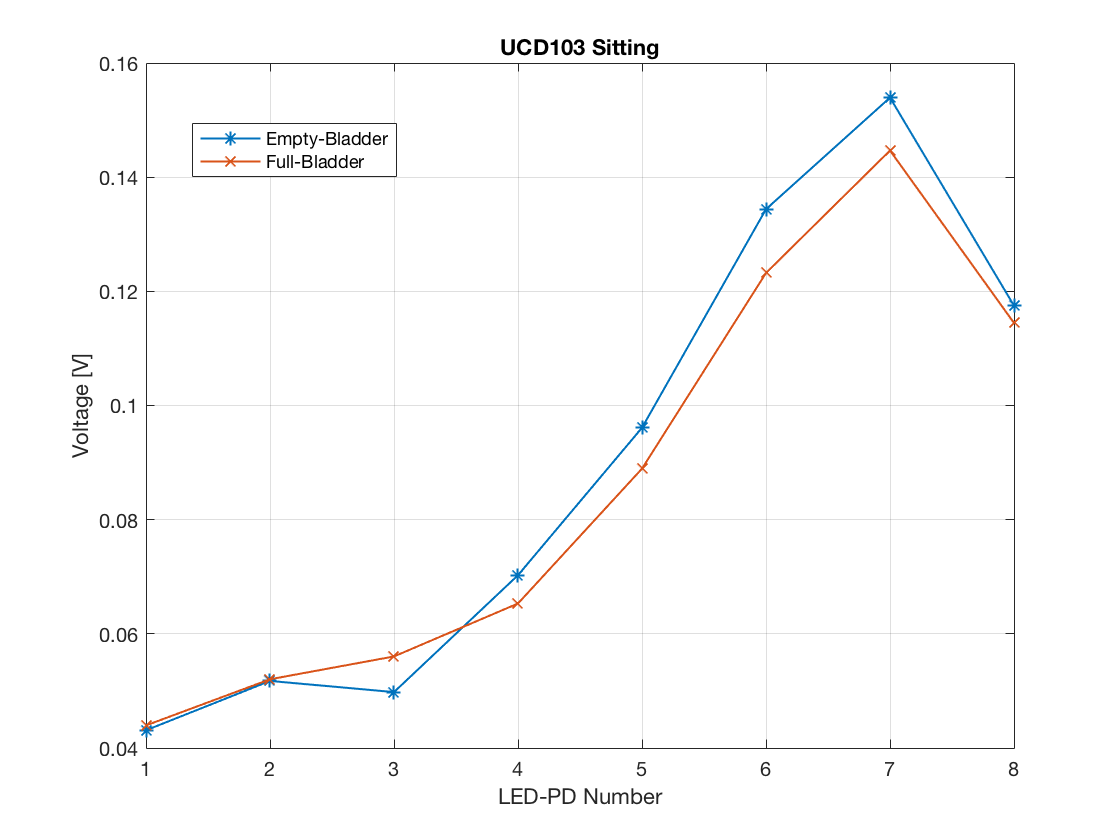}
    \caption{\label{fig:mutli_sitting103}}
  \end{subfigure}
  \begin{subfigure}[a]{0.49\linewidth}
    \centering
    \includegraphics[width=\linewidth]{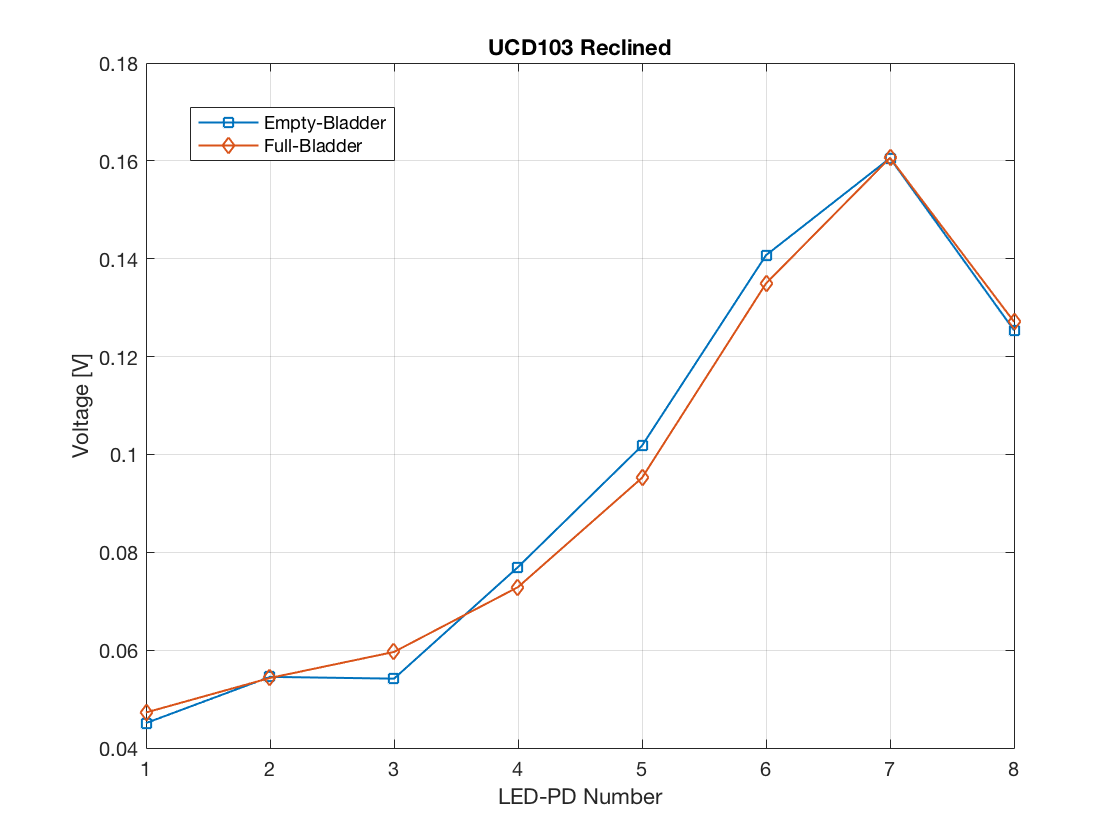}
    \caption{\label{fig:mutli_reclined103}}
  \end{subfigure}%
  \caption[Full and Empty bladder state comparison for UCD103]{Measurements performed on volunteer \textit{UCD103} during the volunteer study. Voltage value reported for each pair is ambient cancelled. Figure \subref{fig:mutli_standing103}), \subref{fig:mutli_sitting103}) and \subref{fig:mutli_reclined103}) show data in standing, sitting and reclined position respectively.}
\label{fig:mutli_103}
\end{figure}

\begin{figure}[H]
\centering%
\begin{subfigure}[a]{0.49\linewidth}
    \centering
    \includegraphics[width=\linewidth]{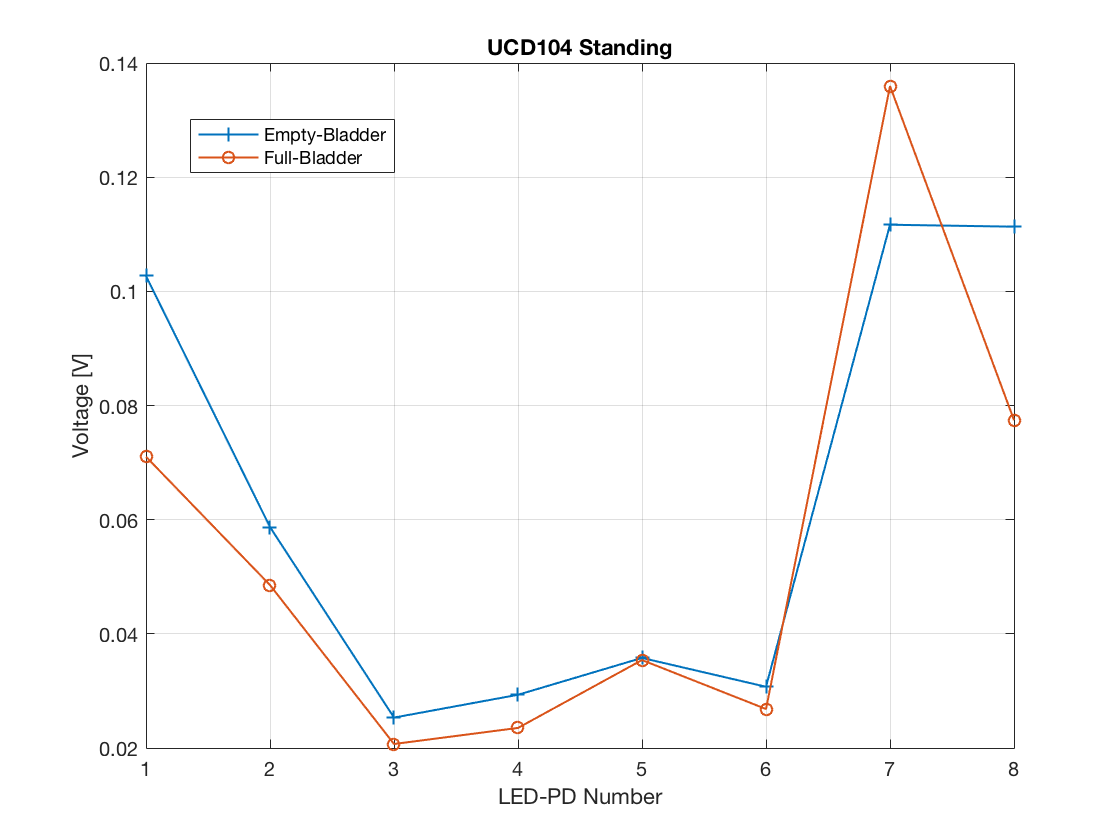}
    \caption{\label{fig:mutli_standing104}}
  \end{subfigure}
  \begin{subfigure}[a]{0.49\linewidth}
    \centering
    \includegraphics[width=\linewidth]{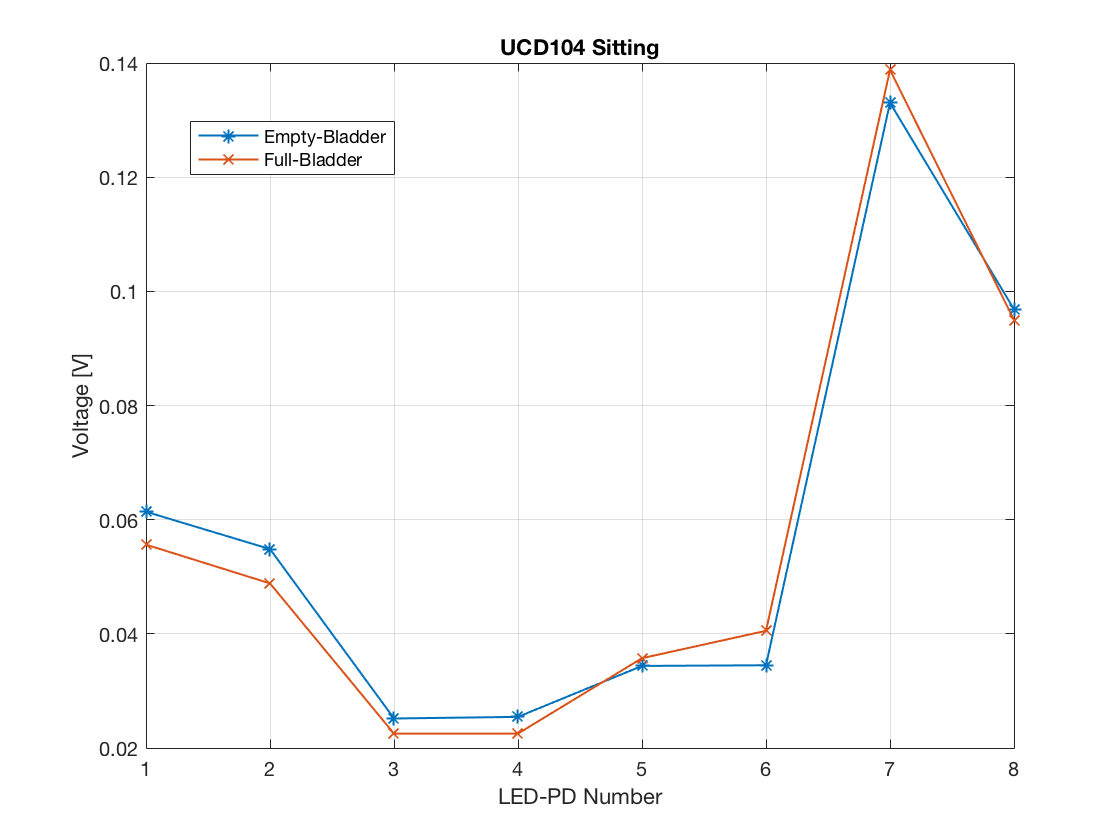}
    \caption{\label{fig:mutli_sitting104}}
  \end{subfigure}
  \begin{subfigure}[a]{0.49\linewidth}
    \centering
    \includegraphics[width=\linewidth]{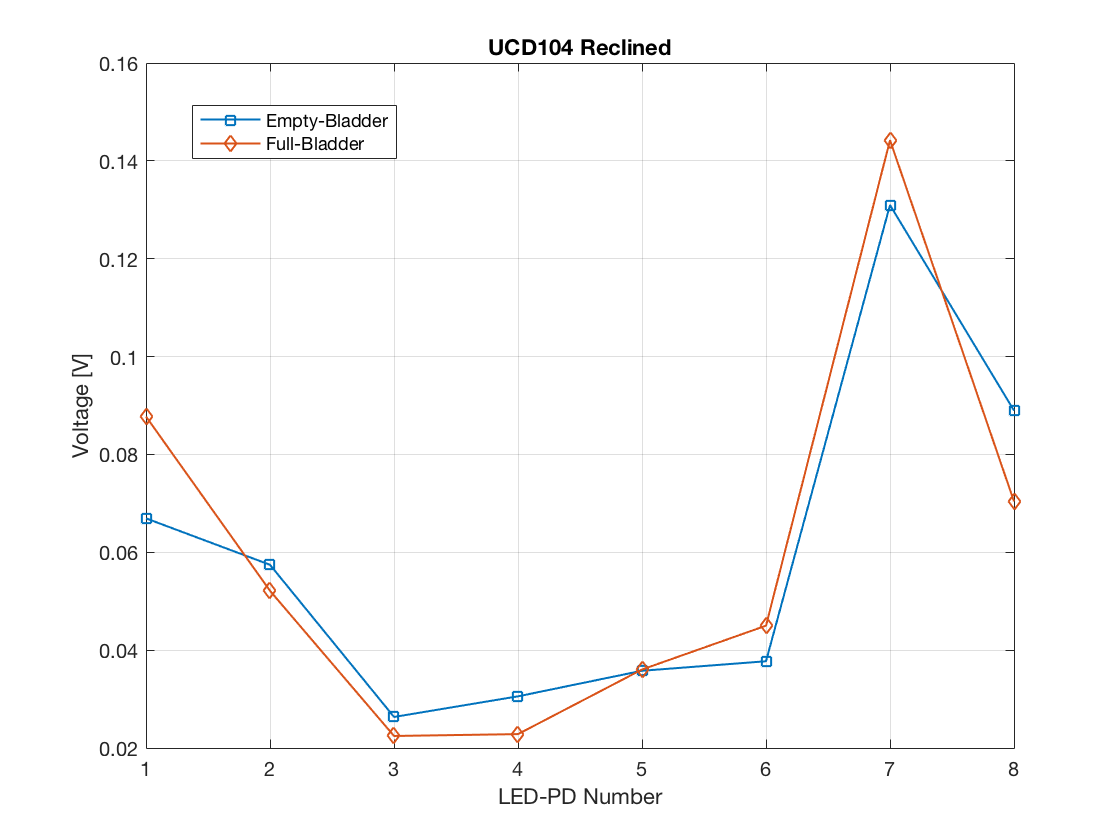}
    \caption{\label{fig:mutli_reclined104}}
  \end{subfigure}%
  \caption[Full and Empty bladder state comparison for UCD104]{Measurements performed on volunteer \textit{UCD104} during the volunteer study. Voltage value reported for each pair is ambient cancelled. Figure \subref{fig:mutli_standing104}), \subref{fig:mutli_sitting104}) and \subref{fig:mutli_reclined104}) show data in standing, sitting and reclined position respectively.}
\label{fig:mutli_104}
\end{figure}

Figure \ref{fig:mutli_103} and \ref{fig:mutli_104} show results from the experiment done on volunteer \textit{UCD103} and \textit{UCD104} in all three aforementioned positions. While the full and empty bladder states in the figures below are represented with orange and blue color, the x-axis values represent optode ID as measurements were performed on each optode pair present on the probe. According to the study done on optical phantoms in section \ref{phantomExperiments}, as volume of liquid inside the bladder increases photons detected at PD decreases because of increased absorption by water.

As seen in figures above, the expected trend is not consistent across optode pairs. Main reason behind this randomness is the variation due to different body shape, size and type, which makes it impossible to co-relate the data between different volunteers. This randomness also makes it difficult to find any underlying patterns by visual inspection. Hence, to test statistical significance of the collected data for full- and empty-bladder states within each optode pair, two-tailed Student's t-test was used. Student's t-test is commonly used statistical test  to check if the two averages (means) are statistically significant and in  two-tailed t-test the critical area of a distribution is two-sided. Table \ref{tab:ttest} shows results across 3 positions for 8 optode pairs present on the probe using data collected from all 10 volunteers. p-value ($p >> 0.01$) across all the data points shows failure to reject the null hypothesis meaning that the collected data is not statistically significant.   

\begin{table}[!ht]
\centering
\caption{Two-tailed Student's t-test on data collected from full and empty bladder state comparison study}
\label{tab:ttest}
\begin{tabular}{|c|c|c|c|}
\hline
& \textbf{Standing} & \textbf{Sitting} & \textbf{Reclined}  \\
\hline
 \textbf{Optode Pair 1}  & 0.587 & 0.232 & 0.409   \\
 \textbf{Optode Pair 2}  & 0.393 & 0.215 & 0.336   \\
 \textbf{Optode Pair 3}  & 0.238 & 0.213 & 0.447  \\
 \textbf{Optode Pair 4}  & 0.104 & 0.852 & 0.144 \\
 \textbf{Optode Pair 5}  & 0.067 & 0.420 & 0.705  \\
 \textbf{Optode Pair 6}  & 0.334 & 0.340 & 0.368  \\
 \textbf{Optode Pair 7}  & 0.331 & 0.264 & 0.329  \\
 \textbf{Optode Pair 8}  & 0.362 & 0.073 & 0.627  \\
 \hline
\end{tabular}
\end{table}

\subsection{Low-frequency longitudinal study with single volunteer}
\label{singlevolunteer}
To remove variation due to different body shape, size and type, device was tested on a single volunteer to check for any visible trend in light intensity at the detector as volume of liquid inside the bladder changes. Also, turning-on the device and probe for a long-period gave insights on its stability and reliability as a function of time. 

Volunteer in this case started with an empty-bladder and readings were taken every 40-minutes in all three aforementioned positions. 200ml of water was consumed by the volunteer after completing each set of readings (readings taken in all three positions for a minute each). The 40 minute gap between sets allowed water to go through volunteer's digestive system and reach his/her bladder. Readings for empty-bladder was taken immediately after voiding and no water was consumed during the \textit{full-} to \textit{empty-bladder} transition. As urine production in human body is an intermittent process, there was no way to quantitatively measure the volume of liquid inside bladder unless it was in \textit{empty-bladder} or \textit{full-bladder} state. Thus, the state of bladder in between these two is simply referred as \textit{partially-filled}. As the volunteer drank water at end of each set, it was assumed that volume of liquid inside the bladder kept increasing, unless \textit{full-bladder} state was reached, post which volunteer voided. Measurements were recorded at each of the optode pairs present on the probe, along with ambient light readings similar to section \ref{multivolunteer} and all sensor values (in volts) reported under the results section are post ambient cancellation. Probe remained attached to the volunteer's abdominal area during the entire length of experiment and was kept in place using self-adhesive wrap to avoid error due to sensor movement.
\subsubsection{Results}
\label{sec:singlevol}

\begin{figure}[H]
\centering%
\begin{subfigure}[a]{0.49\linewidth}
    \centering
    \includegraphics[width=\linewidth]{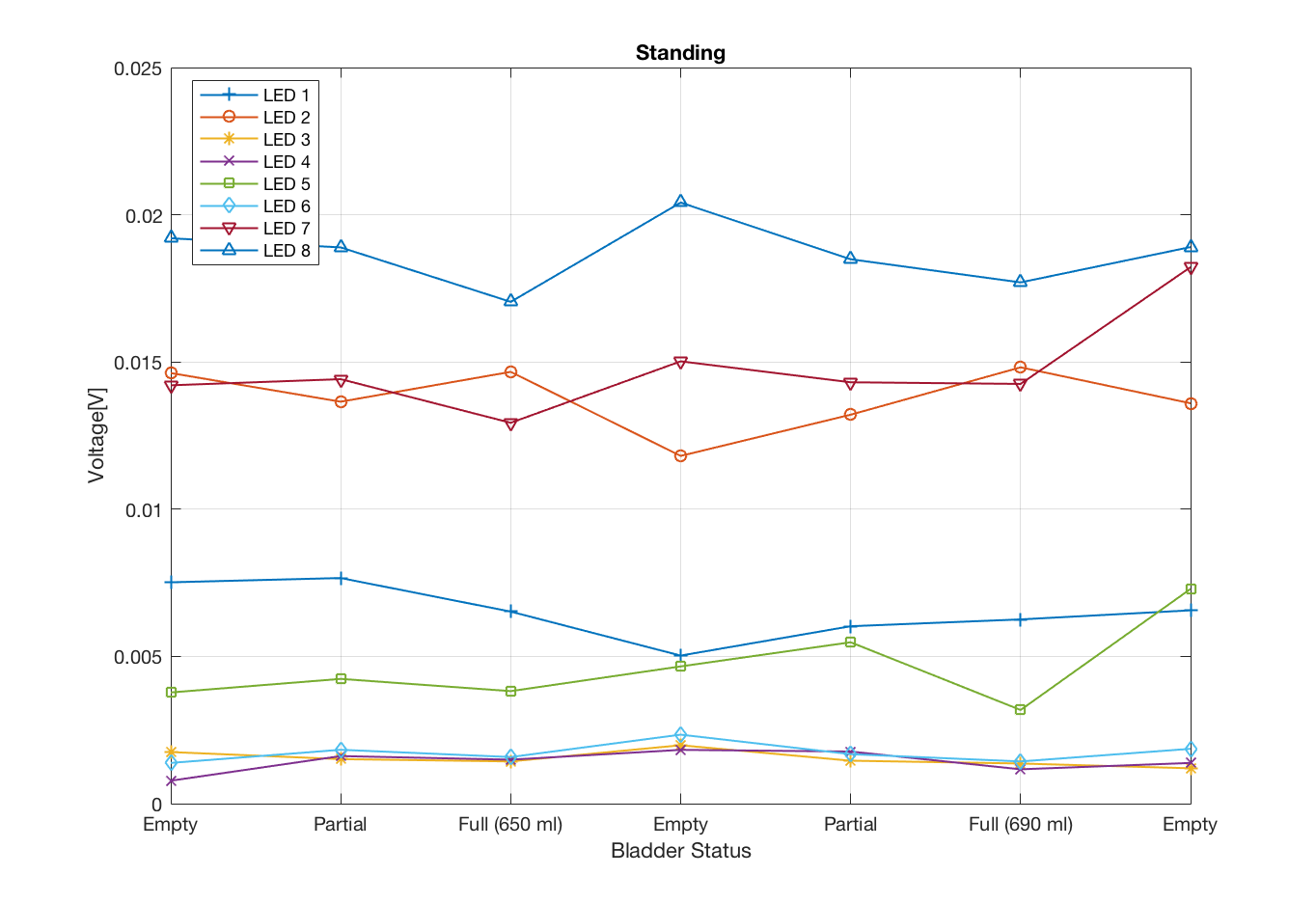}
    \caption{\label{fig:single_standing}}
  \end{subfigure}
  \begin{subfigure}[a]{0.49\linewidth}
    \centering
    \includegraphics[width=\linewidth]{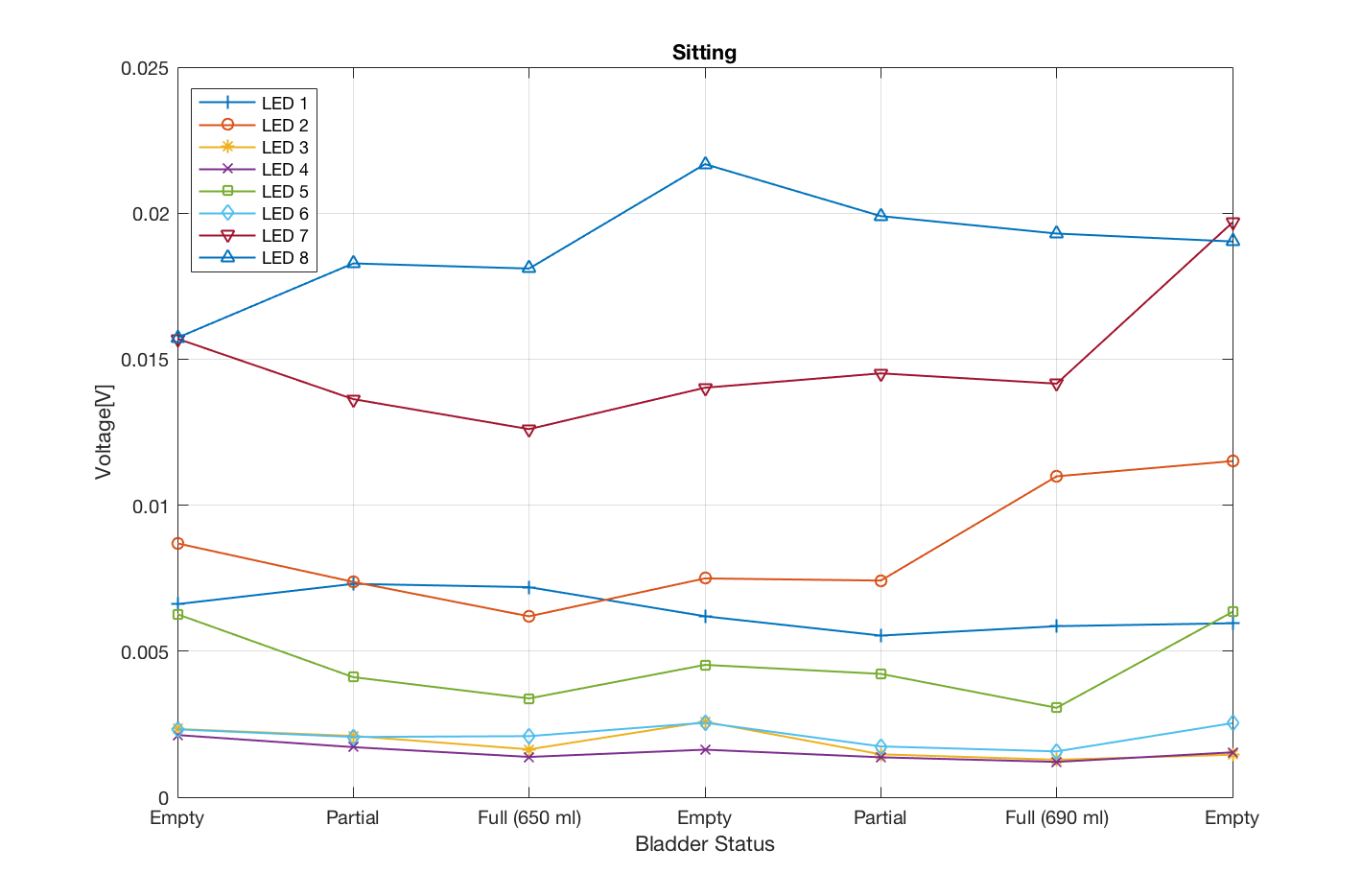}
    \caption{\label{fig:single_sitting}}
  \end{subfigure}
  \begin{subfigure}[a]{0.49\linewidth}
    \centering
    \includegraphics[width=\linewidth]{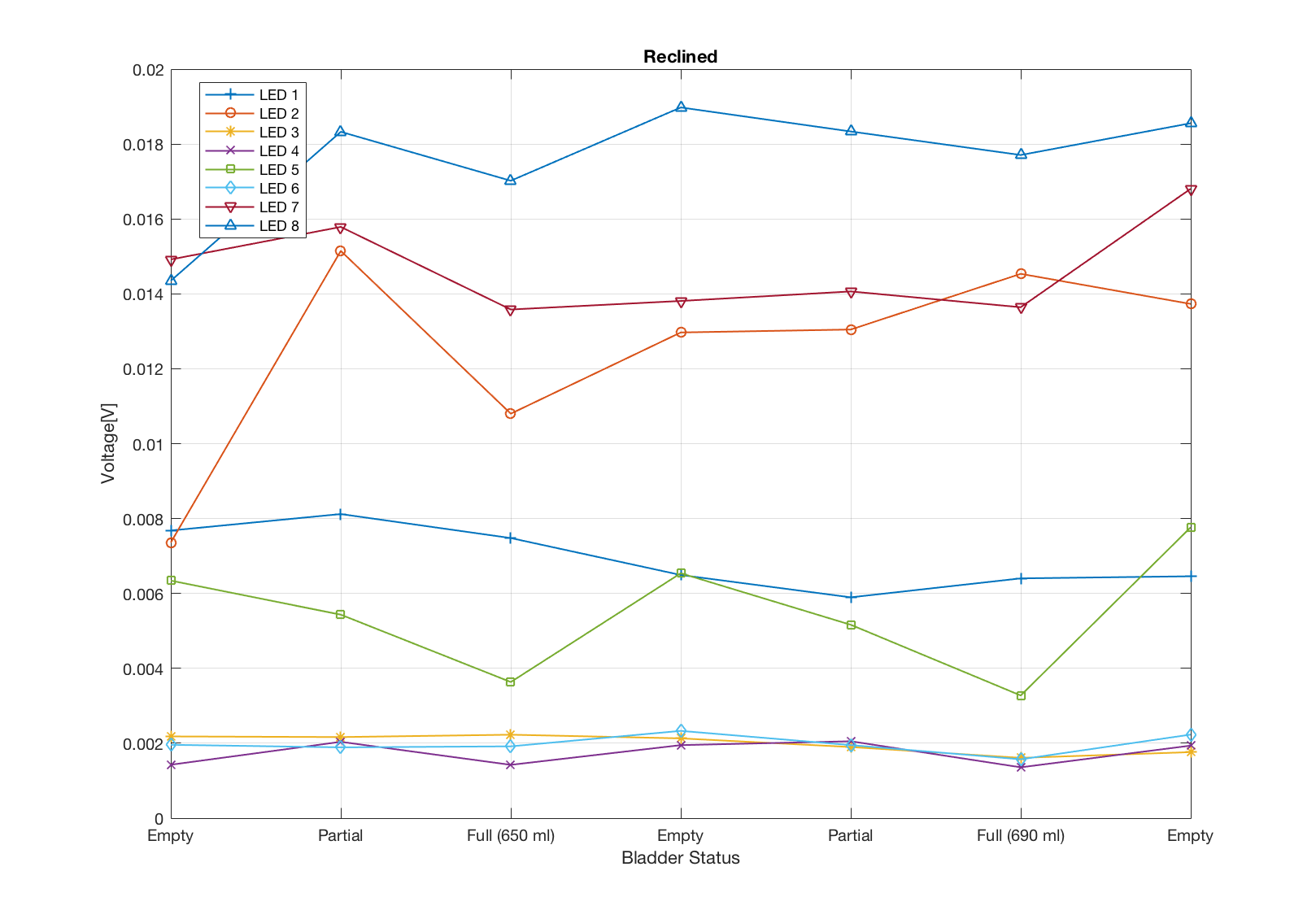}
    \caption{\label{fig:single_reclined}}
  \end{subfigure}%
  \caption[Low-frequency longitudinal study with single volunteer]{Measurements performed on volunteer during the longitudinal study. Bladder status is represented on the x-axis by Empty, Partial or Full state. Figure \subref{fig:single_standing}), \subref{fig:single_sitting}) and \subref{fig:single_reclined}) show the data in standing, sitting and reclined position respectively.}
\label{fig:singlevol}
\end{figure}

Figure \ref{fig:singlevol} shows results from a single volunteer low-frequency longitudinal study in all three positions. It can be seen that optode pairs don't follow a consistent trend, and the random values detected can be attributed to noise. It would thus be useful to quantify the amount of noise that is being detected at the PDs.

\paragraph{Noise Free Measurement}
was done at ADC to measure the amount of noise quantitatively. Calculations were done for each optode pair, keeping gain of ADC as constant. Current measured at the photodiode while LED is on is referred as $I_{Photodiode}$ which is the sum of current due to LED $I_{Pleth}$ and ambient light $I_{Ambient}$. Voltage across the ADC $V_{Diff}$ is governed by the following equation [source:\cite{ti:afe4490}]-

\begin{center}
$V_{Diff} = 2*R_{G}[I_{Pleth}*\frac{R_{F}}{100K} + I_{Ambient}*\frac{R_{F}}{100K} - I_{Cancel}]$ 
\end{center}

where, $R_{F}$ and $R_{G}$ are first and second stage gain control resistances having values 10K$\Omega$ and 100K$\Omega$ respectively and $I_{Cancel}$ is the ambient cancellation current which in this case is 0A. Noise free bits of the ADC $N_{FB}$ can be calculated using the equation [source:\cite{ti:afe4490}]- 

\begin{figure}[h]
    \centering
    \includegraphics[width= 5in]{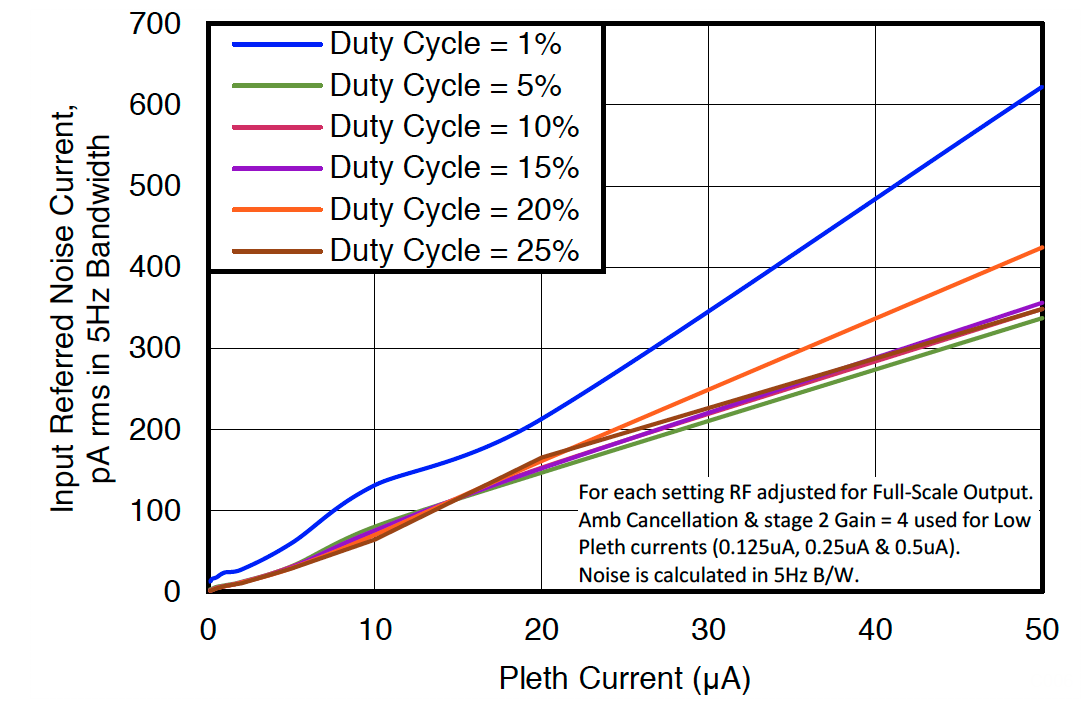}
    \caption[Input-referred RMS noise current.]{Input-referred RMS noise current for the TI AFE4490 \citep{ti:afe4490}.}
    \label{fig:RMSnoise}
\end{figure}

\begin{center}
$N_{FB} = log_{2}[\frac{I_{Photodiode}}{6.6 * I_{Noise}}]$
\end{center}

where, $I_{Noise}$ is the input-referred RMS noise current which can be calculated for corresponding $I_{Pleth}$ value sensed by the PD from the figure \ref{fig:RMSnoise} taken from AFE4490 datasheet. The current setup uses a duty cycle of 25\% with typical values of $I_{Pleth}$ lying in the range of $0.7-1\mu$A. Hence, $I_{Noise}$ value can be approximated as about 10pA. 

Since ADC has a resolution of 22-bits, number of dirty bits $N_{Dirty}$ can be calculated by: $N_{Dirty} = 22 - N_{FB}$, having a typical value of about 12. For the current setup, ADC has a voltage per division value of about 238.42nV and thus, the voltage magnitude of noise $V_{Noise}$ can be calculated using the following equation-

\begin{center}
$V_{Noise} = \frac{238.42nV}{division}* 2^{N_{Dirty}}$ 
\end{center}

SNR values (in dB) can thus be calculated by: $20*log_{10}(\frac{V_{Diff}}{V_{Noise}})$. Table \ref{tab:snr} shows SNR values for each optode pair with measurements taken on the volunteer with an empty bladder in standing position. The SNR values for each optode pair is less than the usual acceptable value of 40dB for embedded systems. As the SD gap for this probe is 4cm, there is a strong possibility of photons not actually hitting the bladder and being received at the detector from shallow depth of field. 

\begin{table}[!ht]
\centering
\caption{Noise Free Measurement for different optode pair having fixed gain settings}
\label{tab:snr}
\begin{tabular}{|c|c|}
\hline
\textbf{Optode Number} & \textbf{SNR value (dB)} \\
\hline
 Optode Pair 1  & 15.28  \\
 Optode Pair 2  & 16.13   \\
 Optode Pair 3  & 3.84  \\
 Optode Pair 4  & 3.62  \\
 Optode Pair 5  & 14.58  \\
 Optode Pair 6  & 3.77   \\
 Optode Pair 7  & 25.92   \\
 Optode Pair 8  & 25.94  \\
 \hline
\end{tabular}
\end{table}

Also, as the probe was attached to the volunteer's abdomen for the entire length of this experiment, it was noticed that probe started to show some signs of deformity and had to be manually brought back into correct shape so that optode pairs were in flush contact with volunteer's skin. This could also have been another potential source of error added to the system during measurements.

\subsection{High-frequency longitudinal study with single volunteer}
\label{highfreq}
Since results from the low-frequency longitudinal study were not very promising and in order to reduce probable error due to volunteer and probe movement as a result of 40 minute gap between readings, measurements were taken in fixed position (sitting) at high frequency to get a better understanding of system behaviour. 

Volunteer, similar to the study covered in section \ref{singlevolunteer}, started with an empty bladder and measurements were recorded at each optode pair present on the probe at high sampling frequency of 1 sample/second (1 sample contains sensor value of all 8LED-PD pairs). Volunteer started with an empty bladder and approximately 100ml of water was consumed every 15min while probe remained attached to the volunteers abdomen during the entire length of study. All sensor values (in volts) reported under the results section are post ambient-cancellation. 

\subsubsection{Results}
\begin{figure}[H]
\centering%
\begin{subfigure}[a]{0.49\linewidth}
    \centering
    \includegraphics[width=\linewidth]{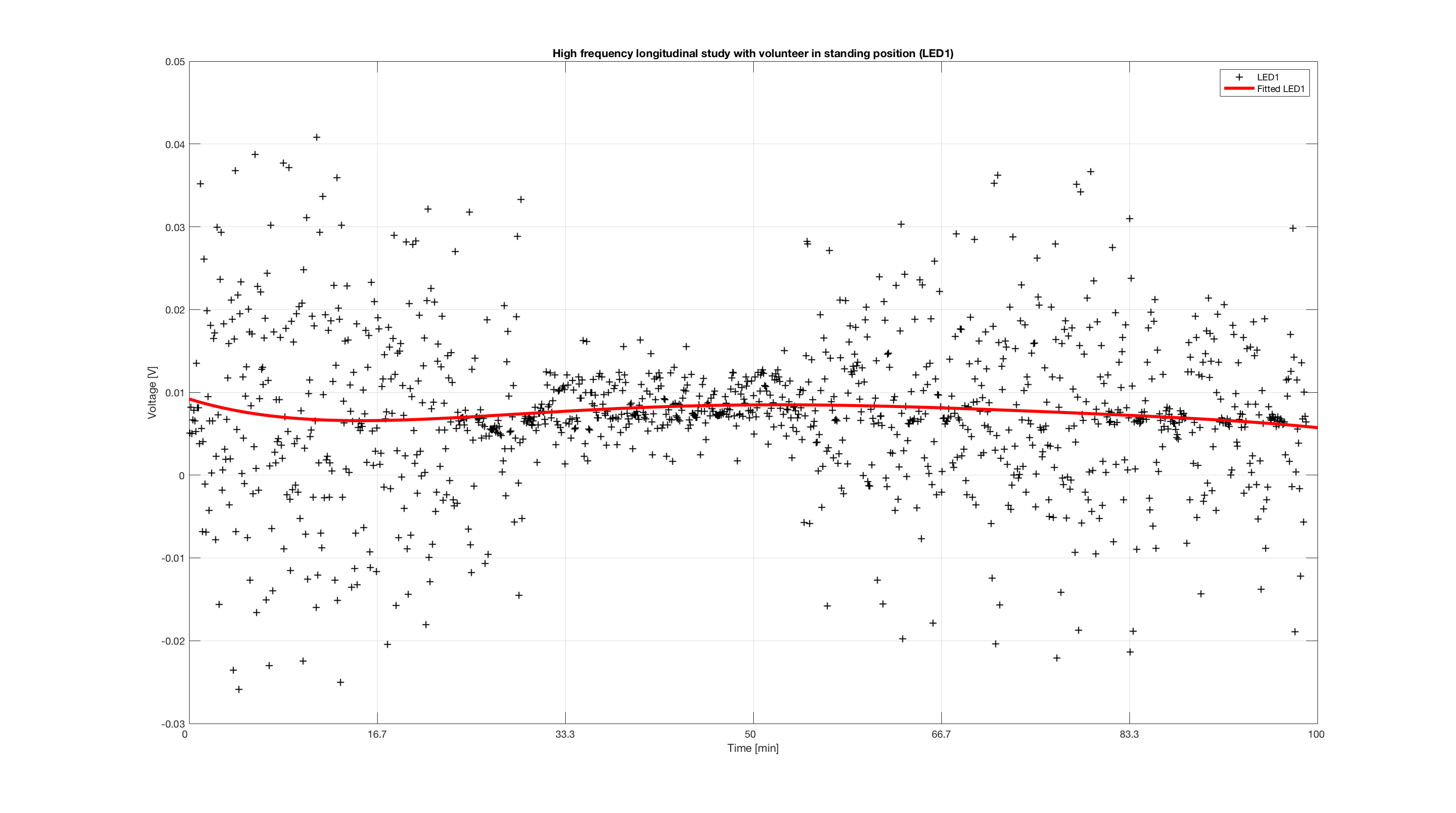}
    \caption{\label{fig:highfreq1}}
  \end{subfigure}
  \begin{subfigure}[a]{0.49\linewidth}
    \centering
    \includegraphics[width=\linewidth]{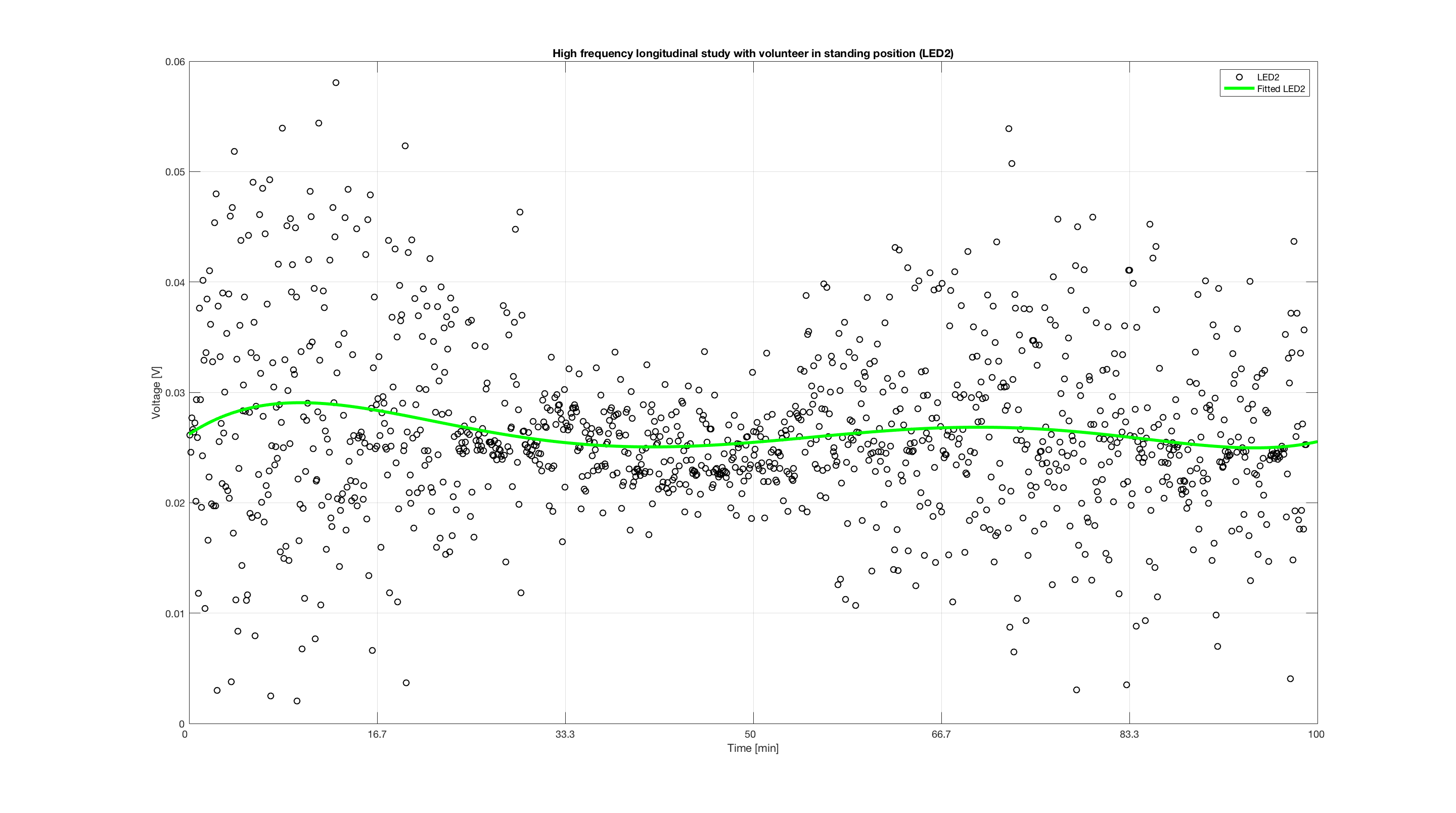}
    \caption{\label{fig:highfreq2}}
  \end{subfigure}
  \begin{subfigure}[a]{0.49\linewidth}
    \centering
    \includegraphics[width=\linewidth]{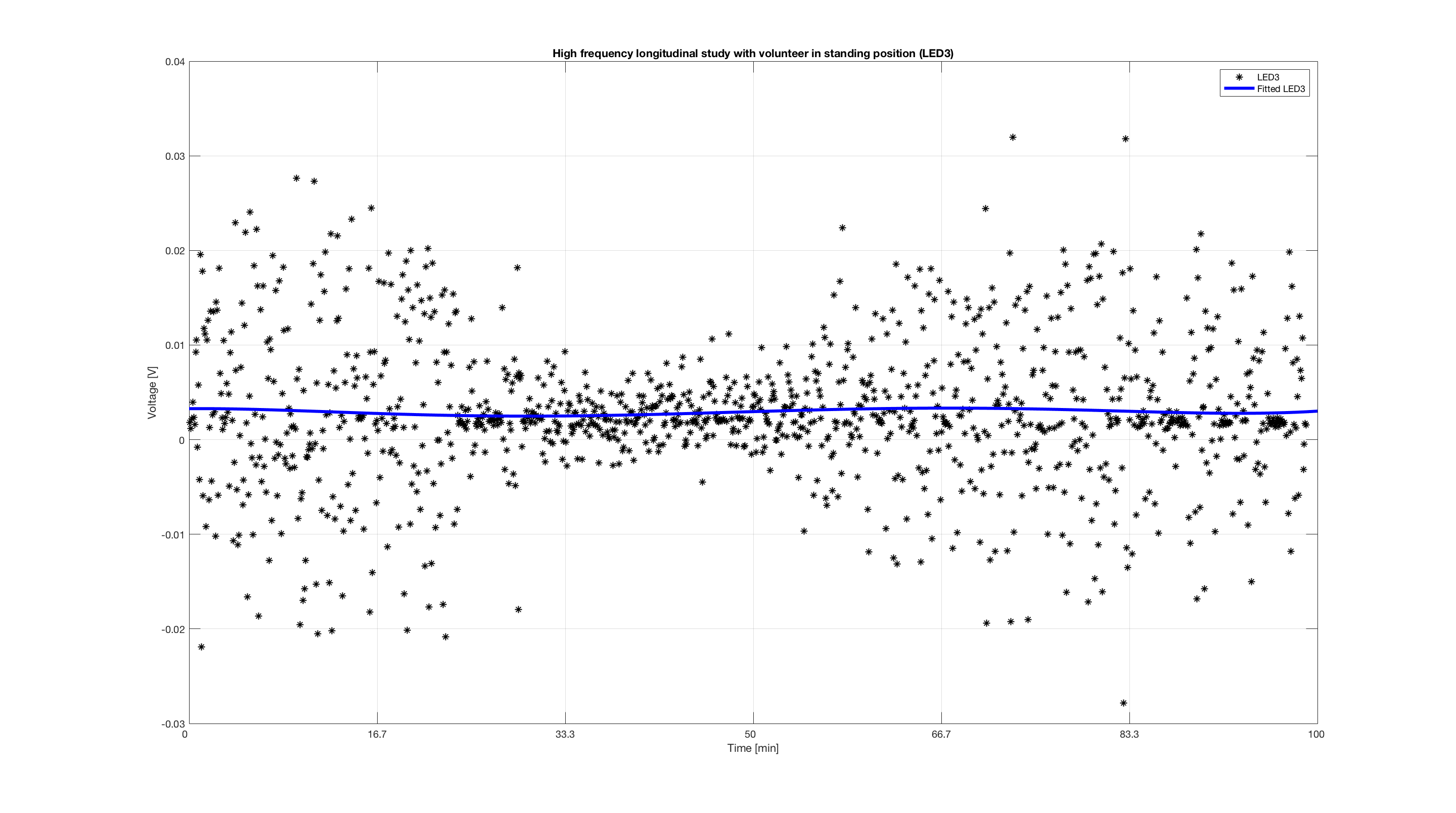}
    \caption{\label{fig:highfreq3}}
    \end{subfigure}
  \begin{subfigure}[a]{0.49\linewidth}
    \centering
    \includegraphics[width=\linewidth]{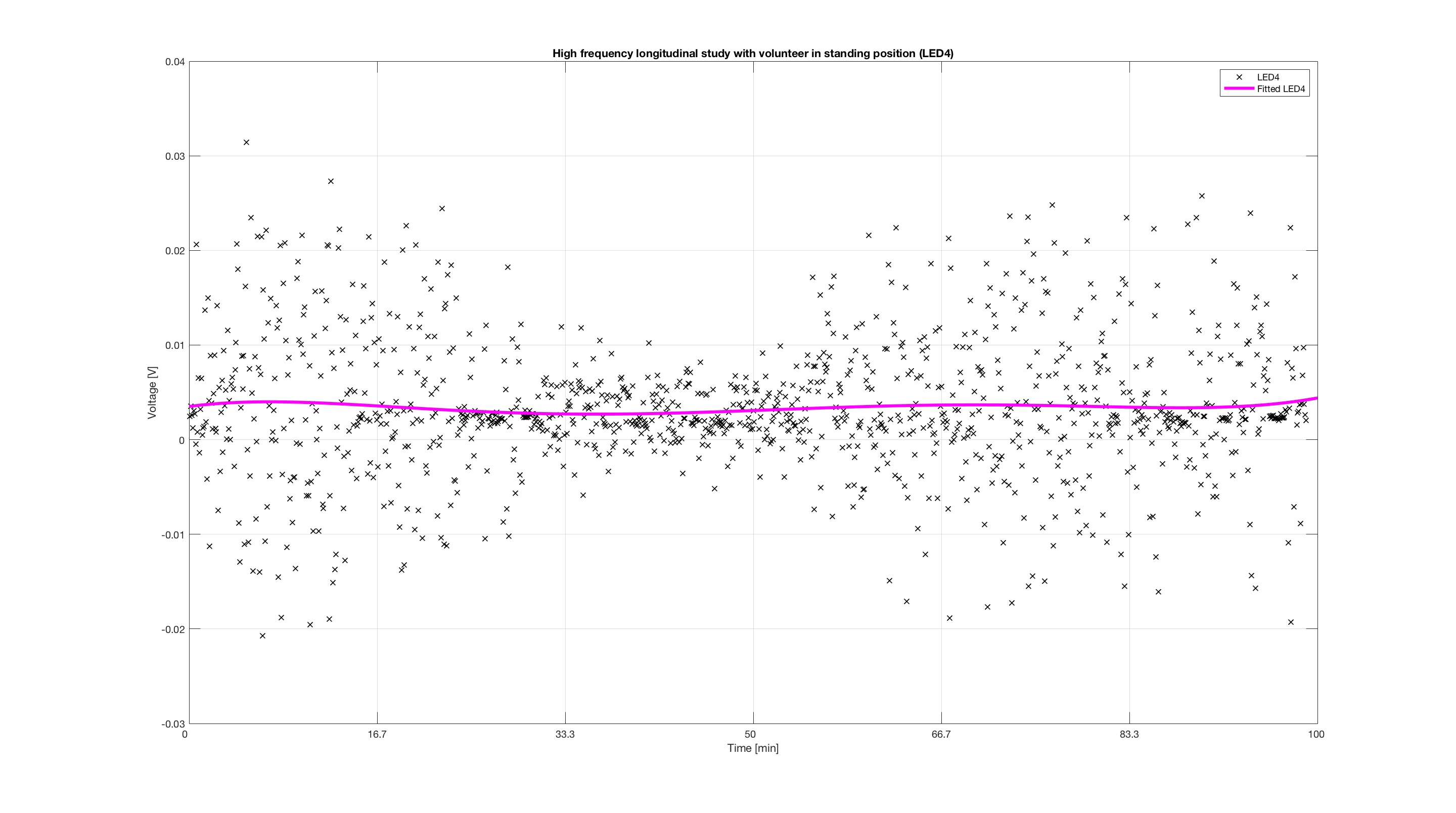}
    \caption{\label{fig:highfreq4}}
  \end{subfigure}
  \caption[High-frequency longitudinal study with single volunteer]{Measurements performed on volunteer during the high-frequency longitudinal study. Each LED data is fitted using a 5$^{th}$ order polynomial function. Figure \subref{fig:highfreq1}) to \subref{fig:highfreq4}) represent data from LED 1 to 4, respectively.  Data from LED 5 to 8 was not reported because of its closeness to Umbilicus.}
\label{fig:highfreq}
\end{figure}
Figure \ref{fig:highfreq} shows result from the single volunteer high-frequency longitudinal study. Data collected from LED 5 to 8 was not reported because of its closeness to Umbilicus. The 5$^{th}$ order polynomial fitted function for each LED helps to ignore small sources of noise and shows fairly stable values. The only variations present are because of volunteer's movement while he/she consumes water.

\begin{figure}[h]
    \centering
    \includegraphics[width= 5in]{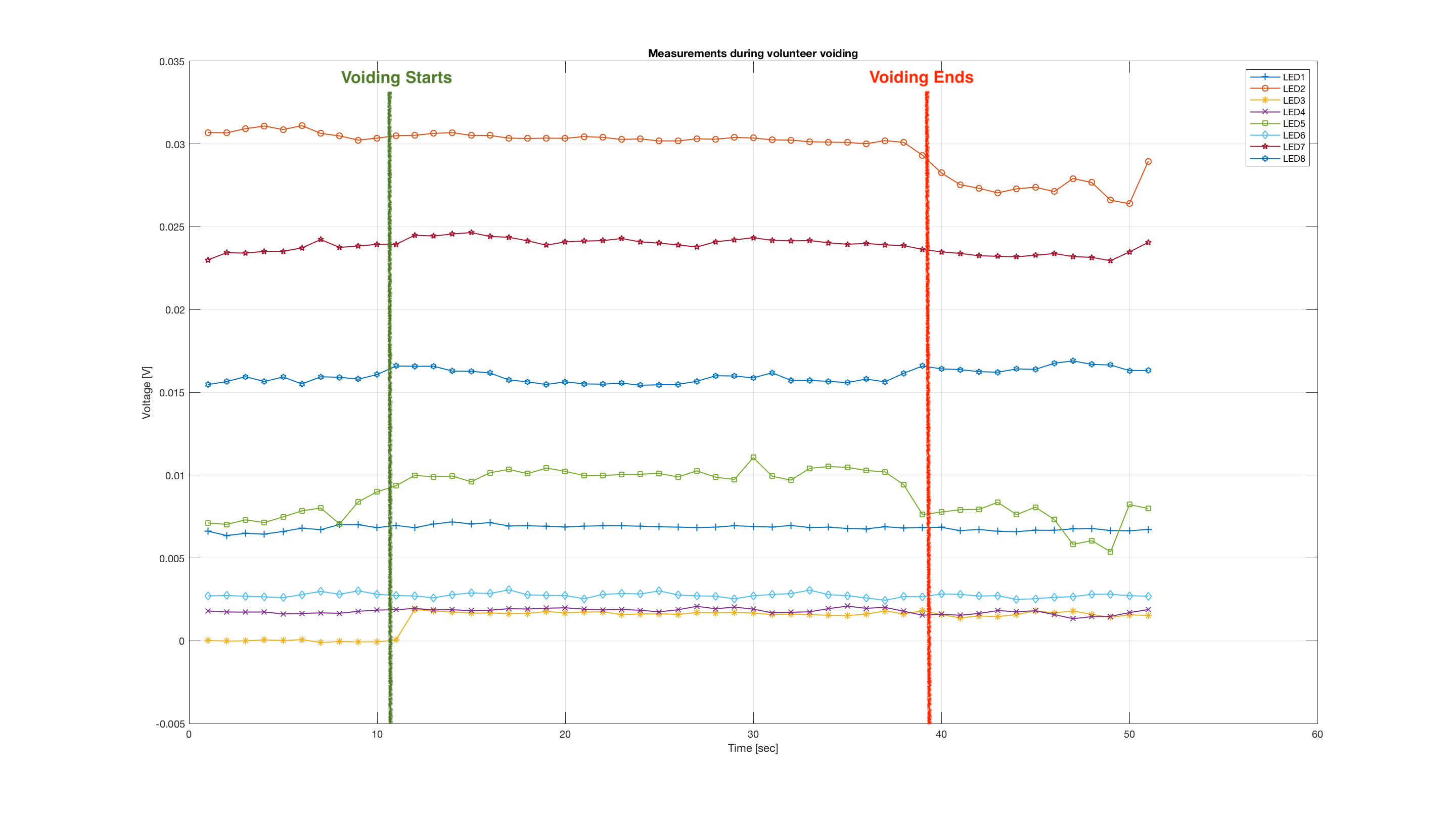}
    \caption[Changes in sensor value during volunteer voiding]{Measurements performed during volunteer voiding. The green line marks the beginning of voiding while red line marks end of voiding.}
    \label{fig:voiding}
\end{figure}

Figure \ref{fig:voiding} shows sensor value as a function of time while volunteer is voiding. Vertical green and red line represents beginning and ending of voiding, respectively. As voiding begins, the urine present in bladder starts to reduce rapidly. The chosen sampling rate is suitable to capture any probable trends in sensor values as the bladder volume decreases. While LED3 shows an increase in sensor value as voiding begins, no clear trends are observed in the measurement which is consistent with other studies done on human subjects.

\subsection{Impact of variable SD distance}
\label{SDimpact}
Goal of this experiment was to visualize penetration depth of photons in the current setup by having variable SD distances on the probe. As mentioned in section \ref{oldOpticalProbe}, depth of photon penetration is approximately half the SD distance and hence, the maximum photon penetration depth can be calculated once the corresponding SD distance is known.  

Since the current probe has eight optode pairs, for getting variable SD distance, instead of recording measurements at each optode pair, one LED was turned on and the measurements were recorded at each of the eight PDs present in the probe. As SD distance increases path length of the photons also increases resulting in higher attenuation. Hence, if value of \textit{LED-on voltage} becomes equal to the \textit{ambient-voltage}, it means that the photons emitted by LED are not actually reaching the detector and detected voltage can be accounted to noise due to ambient light, i.e., in such a case impact of emitter LED on the detector PDs can be considered as zero. 

\subsubsection{Results}
Table \ref{tab:variableSD} shows the impact of variable SD distance on the voltage detected at PDs. For this experiment LED 1 was turned on and the signal was detected on each of the eight PDs. It can be seen that as SD distance increases, value of voltage at detector PD decreases because of the increased path length. Value of ambient-voltage for this measurement was about 3.96E-03V. Hence, the maximum photon penetration depth for the current setup is about 2.25cm which can observed at SD of about 4.5cm.

\begin{table}[!ht]
\centering
\caption{Impact Variable SD distance on the detected signal at PD}
\label{tab:variableSD}
\begin{tabular}{|c|c|c|}
\hline
\textbf{PD Number} & \textbf{SD Distance with } &\textbf{Voltage(V)} \\
& \textbf{respect to LED 1(cm)}&\\
\hline
 PD 1  & 4.0  & 1.61E-02 \\
 PD 2  & 4.5  & 8.81E-03 \\
 PD 3  & 5.7  & 2.64E-03\\
 PD 4  & 7.2  & 3.73E-03\\
 PD 5  & 8.9  & 1.45E-03\\
 PD 6  & 10.8 & 1.27E-03\\
 PD 7  & 12.6 & 6.58E-04\\
 PD 8  & 14.6 & 6.28E-04\\
 \hline
\end{tabular}
\end{table}

\subsection{Impact of lateral photon movement}

In NIRS, photons reaching the PD give information on optical properties of the medium it travels through. In case of air-gap between optodes and tissue structure, some optical signal may leak and travel laterally without entering the tissue structure or may travel at shallow depths through the medium to reach PD. Such photons reaching the PD, contribute to experimental noise as they don't give any information about the tissue medium. Thus, minimizing the impact of lateral photon movement is desired so as to improve the accuracy of the detected signal. 

\begin{figure}[H]
\centering
\begin{subfigure}[a]{0.45\linewidth}
    \centering
    \includegraphics[width=\linewidth]{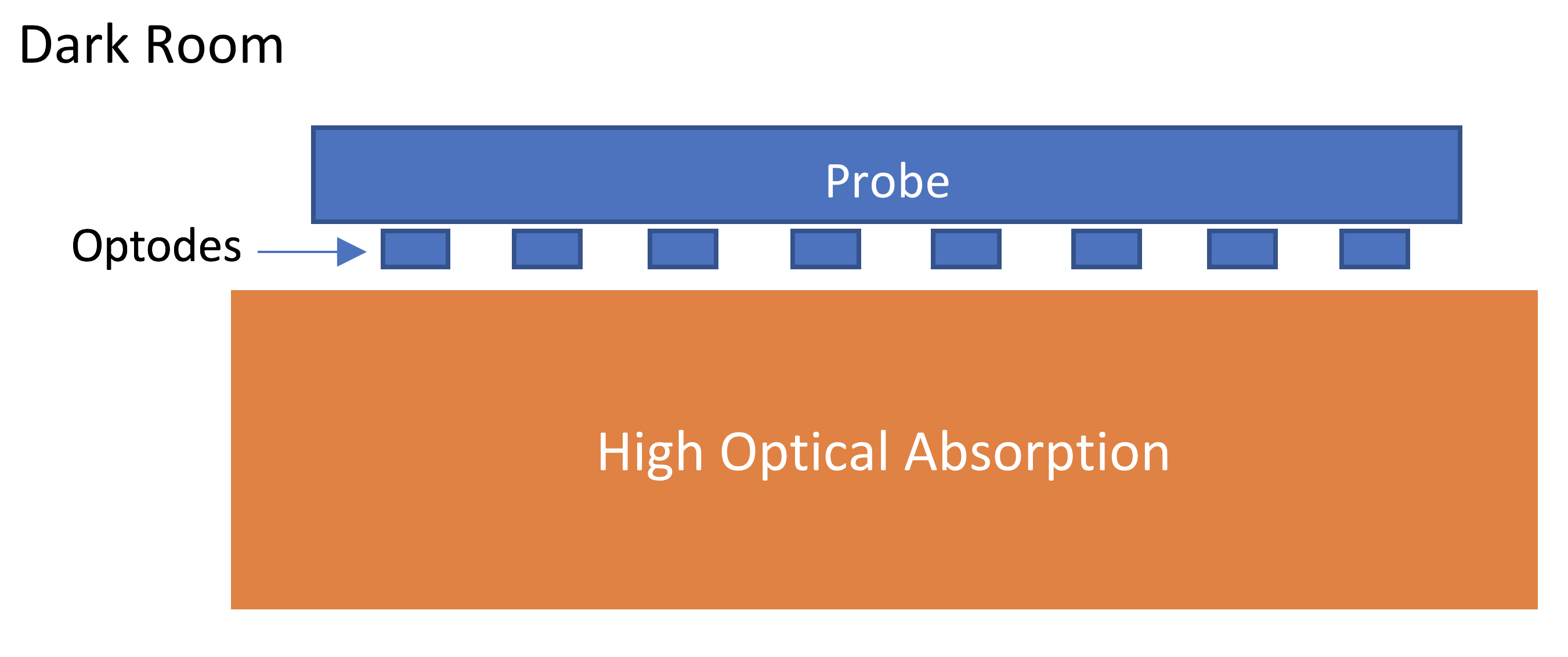}
    \caption{\label{fig:dark1} Side-View}
  \end{subfigure}
  \begin{subfigure}[a]{0.45\linewidth}
    \centering
    \includegraphics[width=\linewidth]{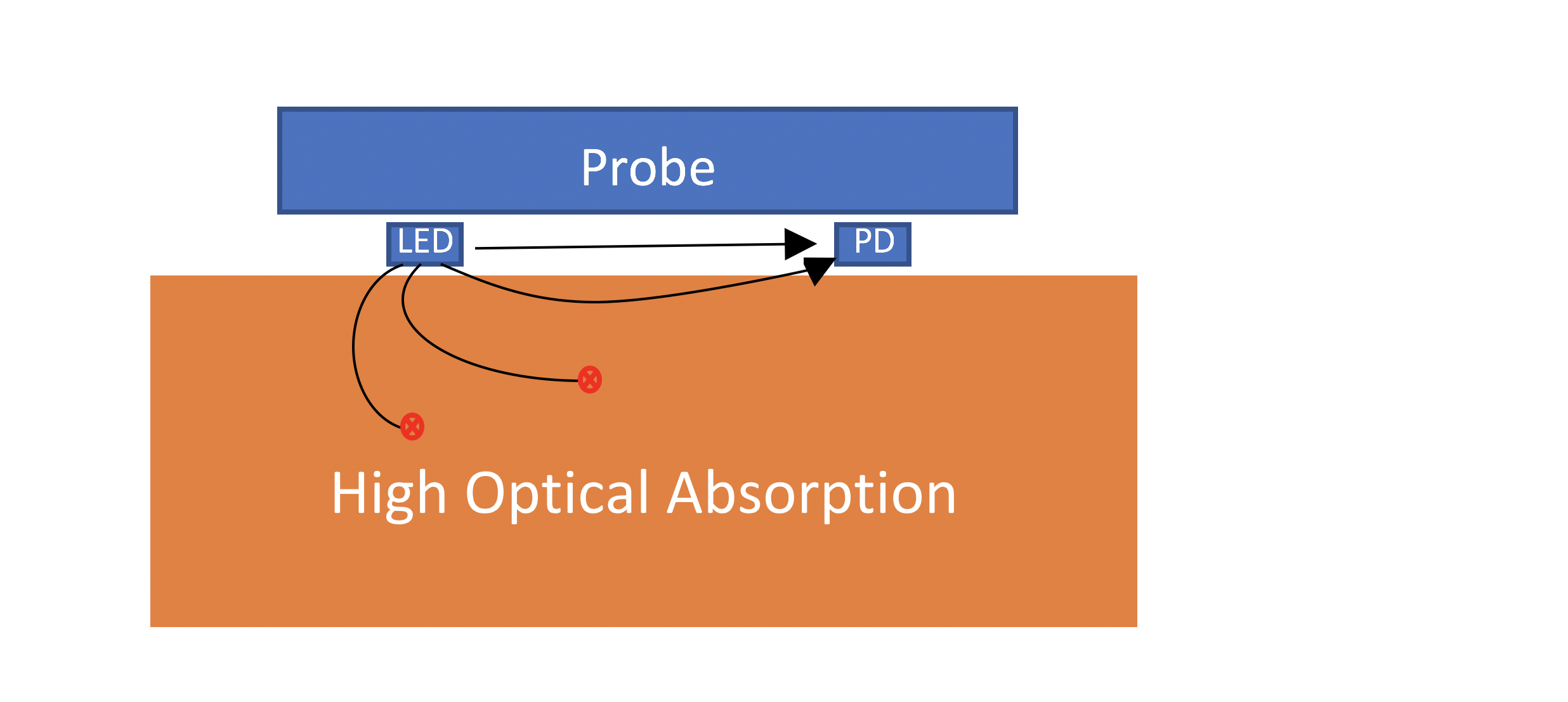}
    \caption{\label{fig:dark2} Front-View}
  \end{subfigure}
  \caption[Setup for measuring impact of lateral photon movement]{\subref{fig:dark1}) Side-view showing the setup, where probe is placed on top of a medium with high optical absorption in a dark room. \subref{fig:dark2}) Front-view showing how photons travel through the system. As medium below the probe has high optical absorption, most photons entering it get absorbed. Thus, photons reaching the PD are as a result of lateral movement or travelling with shallow depths through the medium.}
\label{fig:dark}
\end{figure}

Experimental setup to quantify the impact of lateral photon movement is described in figure \ref{fig:dark}. Block of thick black foam was used as a high optical absorption medium and later the entire setup was placed in a dark room to minimize the impact of ambient light. Thus, most photons entering the medium are absorbed and only the ones travelling laterally or through shallow depths make it to the PDs. The LED-PD measurements were recorded in pairs and the values were reported post ambient cancellation.

Another possible technique to theoretically quantify the impact of lateral photon movement is to check the radiant intensity of LED for viewing angles between  90$^{\circ}$ to about 40$^{\circ}$. The summation of radiant intensities for this viewing angle can thus be approximately considered as causing lateral photon movement and its impact can be calculated using Monte Carlo simulation, described in section \ref{Simulation}. This technique however is not used for calculation of the below mentioned results.

\subsubsection{Results}

Using the setup as described in figure \ref{fig:dark}, the typical voltage value detected as a result of lateral photon movement after multiple runs in a dark room, is about 0.0048V while the typical empty bladder values detected in the human subjects is about 0.035V. Thus, it can be concluded that lateral photon movement contributes to about 14\% of the total detected signal. 

To reduce the impact of lateral photon movement, it is important to have a flushed contact between the optodes and tissue structure. Also, having a thin opaque separation between the LED-PD pairs might help in reducing this noise. 

\begin{comment}
\subsection{Impact of ADC gain on SNR}
\label{gainimpact}
Goal of this experiment is to analyze the impact of varying ADC gain on SNR value of the detected signal.

As mentioned in section \ref{sec:sysarch}, current values measured at the detector PD are sent to ADC which has programmable gain settings. Results from section \ref{sec:singlevol} show low SNR values for single volunteer experiments. Thus, in order to achieve higher SNR values, the gain at transimpedance amplifier is increased by increasing the value of $R_{F}$ while other ADC parameters remained unchanged.

\subsubsection{Results}
Table \ref{tab:ADCgain} shows noise level at ADC as gain increases. It can be seen that SNR value improves as the ADC gain value increases. Even at the highest possible gain setting of ADC with $R_{F} = 1000K\Omega$, the value of SNR is about 32dB which is still less than the usual acceptable value of 40dB for embedded systems. Thus, in order to achieve higher SNR values, system design needs to be changed to accommodate a ADC with higher gain values.

%LED 3 Standing empty
\begin{table}[!ht]
\centering
\caption{Noise Free Measurement with variable ADC gain}
\label{tab:ADCgain}
\begin{tabular}{|c|c|}
\hline
\textbf{$R_{F}$ (K$\Omega$)} & \textbf{SNR value (dB)} \\
\hline
 10.00  & 9.36  \\
 25.00  & 10.91   \\
 50.00  & 11.02  \\
 100.00  & 14.21 \\
 250.00  & 20.95  \\
 500.00  & 26.64 \\
 1000.00  & 32.61  \\
 \hline
\end{tabular}
\end{table}
\end{comment}

\section{Monte Carlo Simulation}
\label{Simulation}
\begin{figure}[h]
    \centering
    \includegraphics[width = 4.5in, height = 4in]{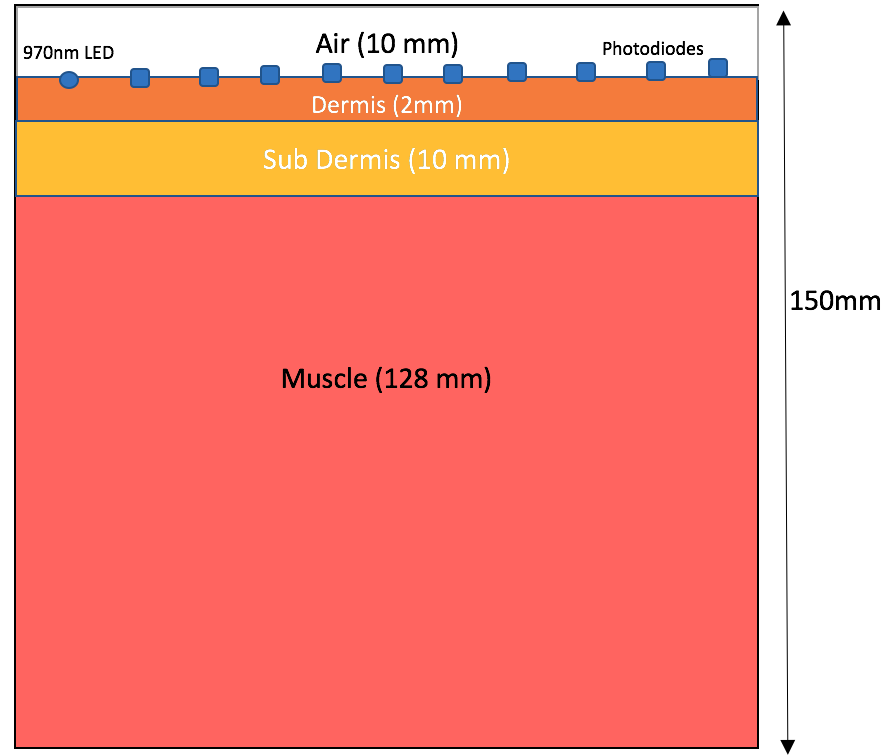}
    \caption[Front-view of the abdomen model used for Monte Carlo Simulations]{Front-view of the abdomen model used for Monte Carlo Simulations. Model is cube with each side being 150mm and having following layers, each of which is also represented by a cube- Air(10mm), Dermis or outer-skin(2mm), Sub-Dermis or subcutaneous fat (10mm) and muscle(128mm). Single 970nm LED source is used and 18 detector PD are placed, each 5mm apart. Blue circle represents LED source while subsequent blue squares represent PDs.}
    \label{fig:simulationModel}
\end{figure}

Monte Carlo Simulation is a numerical simulation method which estimates the light transport in tissue structure and due to stochastic nature of light propagation in tissue, it is widely used for medical purposes \citep{simpson1998near,zhang2007adaptive}. As it can be seen that the trials on human subjects done in section \ref{humanexp} don't show any clear change in sensor values as volume of liquid inside bladder changes. Hence, to explain  inconsistency in results and simulate the behaviour of developed system, a 3-dimensional Monte Carlo algorithm \citep{boas2002three} was used to simulate photon transport in human abdominal tissue structure \citep{mahyasimul}. 3-dimensional tissue model used for simulations can be seen in figure \ref{fig:simulationModel}. Though simulations were done over various SD distances, for purpose of this study only SD gap of 4cm is utilized so as to simulate the current setup. 

As mentioned earlier, 970nm LEDs present on the current probe were driven at 800mA for trials on human subjects. Since LED datasheet gives a radiated power for minimum of 1A of forward current, the curve was extrapolated in order to get a radiated power of 586mW for an input forward current of 800mA \citep{marubeniLED}. As per results from Monte Carlo simulation on model described in figure \ref{fig:simulationModel} for 500million simulated photons \citep{mahyasimul}, radiated optical power of each LED is reduced by an approximate factor of $5.4 \times 10^{-6}$ by the time it reaches detector PD. Hence given 586mW of radiated optical power from the current setup, approximately 3.18$\mu$W of optical power is detected at PD. 

PD having an active area of 0.0625cm$^2$ and responsivity of 0.05mW/cm$^2$ generates a forward current of approximately 2$\mu$A \citep{fairchildPD}. Current generated by PD is fed to an ADC for converting the generated current into voltage values. Considering a single stage ADC gain having $R_{F}$ value 10K$\Omega$, (as used in experiments covered under section \ref{multivolunteer}) it can be seen that the practical and simulated output voltage values are both of the order of ~10mV, taking into account the assumptions made for these calculations. 

\subsection{Results}
Looking at the output voltage values observed by simulation and human experiments, it can be concluded that results from the two studies align with each other. Although model used for simulations does not consider a bladder within the tissue structure, it can be inferred that the developed system described under section \ref{sec:newsysarch}, is probably not able to penetrate tissue structure to actually reach bladder i.e. detected photons are coming from muscle layers after diffused reflectance instead of actually reaching bladder.

Random variation in readings observed during human trials can be explained by various noise sources present inside the human body like abdominal muscle movement, breathing, volunteer movement etc. In order to prove above mentioned analogy, detailed analysis using a more realistic bladder model with different filling states needs to be simulated.

%% file: Chapter5.tex
\chapter{Conclusion and Future Work}
\label{chapter5}
This thesis presented details on various stages during development and testing of a wearable, flexible, non-invasive device for monitoring the amount of urine present inside the human bladder. 

Developed device shows promising results in controlled environments such as optical phantoms, where water present beneath 2cm thick layer of bovine tissue is successfully detected. As volume of water inside the phantom increases, attenuation of photons also increases resulting in decreased signal at the detector PD. Though, results obtained during phantom experiments don't exactly translate to experiments with human subjects but different methodologies tried during the course of this thesis will shed some light in development of next generation bladder volume estimation system.

\section{Future Work}
\label{futurework}
Following are some design considerations worth noting, while developing next generation of probe and optode control system-

\begin{itemize}
\item \textbf{Probe material and size} \\
The current probe used for human trials was made with copper-coated polyimide material which is a flexible bio-material. Wrapping with polyimide tape to prevent copper corrosion makes the probe a little stiff. Another factor which makes flushed contact between the optodes and skin difficult is length of probe (18cm). 

End goal of this study is to have a wearable device in form of an adhesive bio-sensor patch that simply attaches to the users' outer skin.

\item \textbf{Optode Control System} \\
Human trials and simulation results show low current values being detected at PD because of increased signal absorption. Increasing the PD active area can increase the number of detected photons at PD resulting in higher current values.

Also, instead of using a wired connection for off-loading the data and getting user inputs using a wireless-connection like Bluetooth Low-Energy (BLE) would be ideal. This would give more freedom to user as they don't have to be tethered at all times. Eventually, a smart device with built-in Bluetooth like smart watch or smart phone can communicate with the sensor and thus, can be used to notify the user on when is the right time to void.

\item \textbf{Probe attachment} \\
As mentioned earlier, the current probe in long periods of operation, deforms out of shape. Reducing the probe size and making it more flexible will help in improving the optode-skin contact. Also, using a better attachment mechanism will help in keeping probe at a constant position, so that error due to probe movement could be minimized.

\item \textbf{Monte Carlo Simulation} \\
Monte Carlo Simulations are used to estimate light transport in tissue structure and have wide range of  medical applications. Simulation results from section  \ref{Simulation} was used to verify results from human trials and gave insights on the current setup. In a similar way, simulating a more realistic abdominal tissue structure model can give insights on minimum input power required at each LED, expected output current at PD, optimal SD gap etc. which can prove useful for developing next generation bladder volume estimation system. 

\end{itemize}

\subsection{Machine Learning}
Humans are unique individuals and have different shapes, sizes, and body types. This makes it difficult to create a generic system that performs equally well for different individuals when looking at an inherently unique quality such as bladder capacity. The data collected while testing the device on human subjects can be archived over time and thus can later be used to train models using machine learning (ML) for tuning the device which is sensitive to each individual’s characteristics.

For example, data collected from the sensors can be fed to a trained model which using classification algorithms like support vector machines (SVM) and logistic regression would be able to differentiate between full and empty bladder states and give feedback to the user on the right time to void. 

%% file: main.bbl
\begin{thebibliography}{42}
\providecommand{\natexlab}[1]{#1}
\providecommand{\url}[1]{\texttt{#1}}
\expandafter\ifx\csname urlstyle\endcsname\relax
  \providecommand{\doi}[1]{doi: #1}\else
  \providecommand{\doi}{doi: \begingroup \urlstyle{rm}\Url}\fi

\bibitem[Molavi et~al.(2014)Molavi, Shadgan, Macnab, and
  Dumont]{molavi2014noninvasive}
Behnam Molavi, Babak Shadgan, Andrew~J Macnab, and Guy~A Dumont.
\newblock Noninvasive optical monitoring of bladder filling to capacity using a
  wireless near infrared spectroscopy device.
\newblock \emph{IEEE transactions on biomedical circuits and systems},
  8\penalty0 (3):\penalty0 325--333, 2014.

\bibitem[White and Black(2016)]{white2016spinal}
Non-Hispanic White and Non-Hispanic Black.
\newblock Spinal cord injury (sci) facts and figures at a glance.
\newblock \emph{National spinal cord injury statistical center, facts and
  figures at a glance}, 2016.

\bibitem[Broome(2003)]{broome2003impact}
Barbara Ann~Shelton Broome.
\newblock The impact of urinary incontinence on self-efficacy and quality of
  life.
\newblock \emph{Health and Quality of Life Outcomes}, 1\penalty0 (1):\penalty0
  35, 2003.

\bibitem[Robinson and Cardozo(2014)]{robinson2014urinary}
Dudley Robinson and Linda Cardozo.
\newblock Urinary incontinence in the young woman: treatment plans and options
  available.
\newblock \emph{Women’s Health}, 10\penalty0 (2):\penalty0 201--217, 2014.

\bibitem[Gray et~al.(1995)Gray, Rayome, and Anson]{gray1995incontinence}
Mikel Gray, Richard Rayome, and Carol Anson.
\newblock Incontinence and clean intermittent catheterization following spinal
  cord injury.
\newblock \emph{Clinical nursing research}, 4\penalty0 (1):\penalty0 6--18,
  1995.

\bibitem[Nahm et~al.(2015)Nahm, Chen, DeVivo, and Lloyd]{nahm2015bladder}
Laura~S Nahm, Yuying Chen, Michael~J DeVivo, and L~Keith Lloyd.
\newblock Bladder cancer mortality after spinal cord injury over 4 decades.
\newblock \emph{The Journal of urology}, 193\penalty0 (6):\penalty0 1923--1928,
  2015.

\bibitem[Fong et~al.(2018{\natexlab{a}})Fong, Alcantar, Gupta, Kurzrock, and
  Ghiasi]{dfong2018lepsbv}
Daneil Fong, Alejandro~Velazquez Alcantar, Prashant Gupta, Eric Kurzrock, and
  Soheil Ghiasi.
\newblock Non-invasive bladder volume sensing for neurogenic bladder
  dysfunction management.
\newblock \emph{IEEE 15$^{\text{th}}$ International Conference on Wearable and
  Implantable Body Sensor Networks (BSN)}, March 2018{\natexlab{a}}.

\bibitem[Fong et~al.(2018{\natexlab{b}})Fong, Yu, Mao, Saffarpour, Gupta,
  Abueshsheikh, Alcantar, Kurzrock, and Ghiasi]{df2018lepsbv2}
Daneil Fong, Xiaofan Yu, Jiageng Mao, Mahya Saffarpour, Prashant Gupta, Rami
  Abueshsheikh, Alejandro~Velazquez Alcantar, Eric Kurzrock, and Soheil Ghiasi.
\newblock Restoring the sense of bladder fullness for spinal cord injury
  patients.
\newblock \emph{IEEE/ACM 3$^{\text{rd}}$ International Conference on Connected
  Health: Applications, Systems and Engineering Technologies}, September
  2018{\natexlab{b}}.

\bibitem[Jobsis(1977)]{jobsis1977noninvasive}
Frans~F Jobsis.
\newblock Noninvasive, infrared monitoring of cerebral and myocardial oxygen
  sufficiency and circulatory parameters.
\newblock \emph{Science}, 198\penalty0 (4323):\penalty0 1264--1267, 1977.

\bibitem[Scheeren et~al.(2012)Scheeren, Schober, and
  Schwarte]{scheeren2012monitoring}
TWL Scheeren, P~Schober, and LA~Schwarte.
\newblock Monitoring tissue oxygenation by near infrared spectroscopy (nirs):
  background and current applications.
\newblock \emph{Journal of clinical monitoring and computing}, 26\penalty0
  (4):\penalty0 279--287, 2012.

\bibitem[Fong et~al.(2017)Fong, Knoesen, and Ghiasi]{fong2017transabdominal}
Daniel Fong, Andr{\'e} Knoesen, and Soheil Ghiasi.
\newblock Transabdominal fetal pulse oximetry: The case of fetal signal
  optimization.
\newblock In \emph{e-Health Networking, Applications and Services (Healthcom),
  2017 IEEE 19th International Conference on}, pages 1--6. IEEE, 2017.

\bibitem[Pellicer and del Carmen~Bravo(2011)]{pellicer2011near}
Adelina Pellicer and Mar{\'\i}a del Carmen~Bravo.
\newblock Near-infrared spectroscopy: a methodology-focused review.
\newblock In \emph{Seminars in fetal and neonatal medicine}, volume 16, number
  1, pages 42--49. Elsevier, 2011.

\bibitem[Tex(2017)]{ti:afe4490}
\emph{AFE4490 Integrated analog front-end for pulse oximeters}.
\newblock Texas Instruments, August 2017.
\newblock SBAS861.

\bibitem[Kou et~al.(1993)Kou, Labrie, and Chylek]{kou1993refractive}
Linhong Kou, Daniel Labrie, and Petr Chylek.
\newblock Refractive indices of water and ice in the 0.65-to 2.5-$\mu$m
  spectral range.
\newblock \emph{Applied optics}, 32\penalty0 (19):\penalty0 3531--3540, 1993.

\bibitem[Prahl(1999)]{prahl1999tabulated}
SA~Prahl.
\newblock Tabulated molar extinction coefficient for hemoglobin in water.
\newblock \emph{http://omlc. ogi. edu/spectra/hemoglobin/summary. html}, 1999.

\bibitem[van Veen et~al.(2004)van Veen, Sterenborg, Pifferi, Torricelli, and
  Cubeddu]{van2004determination}
Robert~LP van Veen, HJCM Sterenborg, A~Pifferi, A~Torricelli, and R~Cubeddu.
\newblock Determination of vis-nir absorption coefficients of mammalian fat,
  with time-and spatially resolved diffuse reflectance and transmission
  spectroscopy.
\newblock In \emph{Biomedical Topical Meeting}, page SF4. Optical Society of
  America, 2004.

\bibitem[Wilson et~al.(2015)Wilson, Nadeau, Jaworski, Tromberg, and
  Durkin]{wilson2015review}
Robert~H Wilson, Kyle~P Nadeau, Frank~B Jaworski, Bruce~J Tromberg, and
  Anthony~J Durkin.
\newblock Review of short-wave infrared spectroscopy and imaging methods for
  biological tissue characterization.
\newblock \emph{Journal of biomedical optics}, 20\penalty0 (3):\penalty0
  030901, 2015.

\bibitem[College(2013)]{openstax}
OpenStax College.
\newblock \emph{Anatomy \& Physiology}.
\newblock OpenStax College, 2013.

\bibitem[Rose et~al.(2015)Rose, Parker, Jefferson, and
  Cartmell]{rose2015characterization}
C~Rose, Alison Parker, Bruce Jefferson, and Elise Cartmell.
\newblock The characterization of feces and urine: a review of the literature
  to inform advanced treatment technology.
\newblock \emph{Critical reviews in environmental science and technology},
  45\penalty0 (17):\penalty0 1827--1879, 2015.

\bibitem[Kristiansen et~al.(2004{\natexlab{a}})Kristiansen, Ringgaard, Nygaard,
  and Djurhuus]{kristiansen2004effect}
Niels~Kristian Kristiansen, Steffen Ringgaard, Hans Nygaard, and Jens~Christian
  Djurhuus.
\newblock Effect of bladder volume, gender and body position on the shape and
  position of the urinary bladder.
\newblock \emph{Scandinavian journal of urology and nephrology}, 38\penalty0
  (6):\penalty0 462--468, 2004{\natexlab{a}}.

\bibitem[Patel and Rickards(2010)]{patel2010imaging}
Uday Patel and David Rickards.
\newblock \emph{Imaging and urodynamics of the lower urinary tract}.
\newblock Springer, 2010.

\bibitem[Dreher et~al.(1972)Dreher, Timm, and Bradley]{dreher1972bladder}
Robert~D Dreher, Gerald~W Timm, and William~E Bradley.
\newblock Bladder volume sensing by local distension measurement.
\newblock \emph{IEEE Transactions on Biomedical Engineering}, BME-19\penalty0
  (3):\penalty0 247--248, 1972.

\bibitem[Wang et~al.(2009)Wang, Hou, Zheng, Zhang, Chen, and
  Xu]{wang2009design}
Jianhuo Wang, Chunlin Hou, Xianyou Zheng, Wei Zhang, Aimin Chen, and Zhen Xu.
\newblock Design and evaluation of a new bladder volume monitor.
\newblock \emph{Archives of physical medicine and rehabilitation}, 90\penalty0
  (11):\penalty0 1944--1947, 2009.

\bibitem[Coosemans and Puers(2005)]{coosemans2005autonomous}
Johan Coosemans and Robert Puers.
\newblock An autonomous bladder pressure monitoring system.
\newblock \emph{Sensors and actuators A: Physical}, 123:\penalty0 155--161,
  2005.

\bibitem[Majerus et~al.(2016)Majerus, Basu, Makovey, Wang, Zhui, Zorman, Ko,
  and Damaser]{majerus2016wireless}
Steve Majerus, Anisha~S Basu, Iryna Makovey, Peng Wang, Hui Zhui, Christian
  Zorman, Wen Ko, and Margot~S Damaser.
\newblock Wireless bladder pressure monitor for closed-loop bladder
  neuromodulation.
\newblock In \emph{SENSORS, 2016 IEEE}, pages 1--3. IEEE, 2016.

\bibitem[Chen et~al.(2015{\natexlab{a}})Chen, Hsieh, Fan, Lai, Chen, Wei, and
  Peng]{chen2015design}
Shih-Ching Chen, Tsung-Hsun Hsieh, Wen-Jia Fan, Chien-Hung Lai, Chun-Lung Chen,
  Wei-Feng Wei, and Chih-Wei Peng.
\newblock Design and evaluation of potentiometric principles for bladder volume
  monitoring: a preliminary study.
\newblock \emph{Sensors}, 15\penalty0 (6):\penalty0 12802--12815,
  2015{\natexlab{a}}.

\bibitem[Schlebusch et~al.(2014)Schlebusch, Nienke, Leonhardt, and
  Walter]{schlebusch2014bladder}
Thomas Schlebusch, Steffen Nienke, S~Leonhardt, and M~Walter.
\newblock Bladder volume estimation from electrical impedance tomography.
\newblock \emph{Physiological measurement}, 35\penalty0 (9):\penalty0 1813,
  2014.

\bibitem[Macnab et~al.(2005)Macnab, Gagnon, and Stothers]{macnab2005clinical}
AJ~Macnab, RE~Gagnon, and L~Stothers.
\newblock Clinical nirs of the urinary bladder--a demonstration case report.
\newblock \emph{Journal of Spectroscopy}, 19\penalty0 (4):\penalty0 207--212,
  2005.

\bibitem[Kiely et~al.(1987)Kiely, Hartnell, Gibson, and
  Williams]{kiely1987measurement}
EA~Kiely, GG~Hartnell, RN~Gibson, and Gordon Williams.
\newblock Measurement of bladder volume by real-time ultrasound.
\newblock \emph{BJU International}, 60\penalty0 (1):\penalty0 33--35, 1987.

\bibitem[Dicuio et~al.(2005)Dicuio, Pomara, Fabris, Ales, Dahlstrand, and
  Morelli]{dicuio2005measurements}
Mauro Dicuio, Giorgio Pomara, F~Menchini Fabris, Valeria Ales, Christer
  Dahlstrand, and Girolamo Morelli.
\newblock Measurements of urinary bladder volume: comparison of five ultrasound
  calculation methods in volunteers.
\newblock \emph{Arch Ital Urol Androl}, 77\penalty0 (1):\penalty0 60--62, 2005.

\bibitem[Kristiansen et~al.(2004{\natexlab{b}})Kristiansen, Djurhuus, and
  Nygaard]{kristiansen2004design}
NK~Kristiansen, Jens~Christian Djurhuus, and Hans Nygaard.
\newblock Design and evaluation of an ultrasound-based bladder volume monitor.
\newblock \emph{Medical and Biological Engineering and Computing}, 42\penalty0
  (6):\penalty0 762--769, 2004{\natexlab{b}}.

\bibitem[Van~der Zee et~al.(1992)Van~der Zee, Cope, Arridge, Essenpreis,
  Potter, Edwards, Wyatt, McCormick, Roth, Reynolds,
  et~al.]{van1992experimentally}
P~Van~der Zee, M~Cope, SR~Arridge, M~Essenpreis, LA~Potter, AD~Edwards,
  JS~Wyatt, DC~McCormick, SC~Roth, EOR Reynolds, et~al.
\newblock Experimentally measured optical pathlengths for the adult head, calf
  and forearm and the head of the newborn infant as a function of inter optode
  spacing.
\newblock In \emph{Oxygen transport to tissue XIII}, pages 143--153. Springer,
  1992.

\bibitem[Zonios and Dimou(2006)]{zonios2006modeling}
George Zonios and Aikaterini Dimou.
\newblock Modeling diffuse reflectance from semi-infinite turbid media:
  application to the study of skin optical properties.
\newblock \emph{Optics express}, 14\penalty0 (19):\penalty0 8661--8674, 2006.

\bibitem[Chen et~al.(2015{\natexlab{b}})Chen, Lu, Chen, and
  Feng]{chen2015breathable}
Ying Chen, Bingwei Lu, Yihao Chen, and Xue Feng.
\newblock Breathable and stretchable temperature sensors inspired by skin.
\newblock \emph{Scientific reports}, 5:\penalty0 11505, 2015{\natexlab{b}}.

\bibitem[DIN(2011)]{din60601}
EN~DIN.
\newblock 60601-2-57.
\newblock Technical report, VDE 0750-2-57: 2011-11, Medical electrical
  equipment—Part 2--57: Particular requirements for the basic safety and
  essential performance of non-laser light source equipment intended for
  therapeutic, diagnostic, monitoring and cosmetic/aesthetic use IEC
  60601-2-57, 2011.

\bibitem[Commission et~al.(2006)]{international2006iec}
International~Electrotechnical Commission et~al.
\newblock {IEC} 62471: 2006.
\newblock \emph{Photobiological safety of lamps and lamp systems. Geneva: IEC},
  2006.

\bibitem[Simpson et~al.(1998)Simpson, Kohl, Essenpreis, and
  Cope]{simpson1998near}
C~Rebecca Simpson, Matthias Kohl, Matthias Essenpreis, and Mark Cope.
\newblock Near-infrared optical properties of ex vivo human skin and
  subcutaneous tissues measured using the monte carlo inversion technique.
\newblock \emph{Physics in Medicine \& Biology}, 43\penalty0 (9):\penalty0
  2465, 1998.

\bibitem[Zhang et~al.(2007)Zhang, Brown, and Strangman]{zhang2007adaptive}
Quan Zhang, Emery~N Brown, and Gary~E Strangman.
\newblock Adaptive filtering for global interference cancellation and real-time
  recovery of evoked brain activity: a monte carlo simulation study.
\newblock \emph{Journal of biomedical optics}, 12\penalty0 (4):\penalty0
  044014, 2007.

\bibitem[Boas et~al.(2002)Boas, Culver, Stott, and Dunn]{boas2002three}
David~A Boas, JP~Culver, JJ~Stott, and AK~Dunn.
\newblock Three dimensional monte carlo code for photon migration through
  complex heterogeneous media including the adult human head.
\newblock \emph{Optics express}, 10\penalty0 (3):\penalty0 159--170, 2002.

\bibitem[Saffarpour and Ghiasi(2018)]{mahyasimul}
Mahya Saffarpour and Soheil Ghiasi.
\newblock A design space exploration {(DSE)} on non-invasive sensing of bladder
  filling using near infrared spectroscopy {(NIRS)}.
\newblock \emph{ArXiv e-prints}, June 2018.

\bibitem[Mar(2014)]{marubeniLED}
\emph{Marubeni SMBB970D-1100-02, High Power LED.}
\newblock Marubeni, 2014.
\newblock
  \url{http://tech-led.com/wp-content/uploads/2017/09/SMBB970D-1100-02.pdf}.

\bibitem[Fai(2016)]{fairchildPD}
\emph{QSB34GR/QSB34ZR/QSB34CGR/QSB34CZR Surface-Mount Silicon Pin Photodiode}.
\newblock Fairchild, September 2016.
\newblock \url{https://www.mouser.com/ds/2/149/QSB34-1011917.pdf}.

\end{thebibliography}
